\documentclass[fleqn,usenatbib]{mnras}

\usepackage{newtxtext,newtxmath}

\usepackage[T1]{fontenc}

\DeclareRobustCommand{\VAN}[3]{#2}
\let\VANthebibliography\thebibliography
\def\thebibliography{\DeclareRobustCommand{\VAN}[3]{##3}\VANthebibliography}

\usepackage{graphicx}	
\usepackage{amsmath}	
\usepackage{tabularx}      
\usepackage[separate-uncertainty=true]{siunitx} 
\usepackage[normalem]{ulem} 
\usepackage{orcidlink}

\DeclareSIUnit\year{yr}

\renewcommand\apjl{ApJL} 
\newcommand\natastr{Nature Astron.} 
\newcommand\natcom{Nature Comm.} 
\newcommand\ptrsa{Philos. Trans. Royal Soc. A: Math., Phys. \& Eng. Sci.} 
\newcommand\sciadv{Science Adv.} 
\newcommand\acsESC{ACS Earth \& Sp. Chem.} 
\newcommand\ijms{Int. J. Mass Spec.} 
\newcommand\plansj{Plan. Sci. J.} 
\newcommand\emp{Earth, Moon, \& Plan.} 
\newcommand\pccp{Phys. Chem. Chem. Phys.}      

\title[Volatiles in 67P]{Volatiles in the H$_2$O and CO$_2$ ices of comet 67P/Churyumov-Gerasimenko}

\author[Rubin et al.]{Martin Rubin\,\textsuperscript{\orcidlink{0000-0001-6549-3318}},$^{1}$\thanks{E-mail: martin.rubin@unibe.ch}
Kathrin Altwegg\,\textsuperscript{\orcidlink{0000-0002-2677-8238}},$^{1}$
Jean-Jacques Berthelier\,\textsuperscript{\orcidlink{0000-0002-4261-2772}}$,^{2}$
Michael R. Combi\,\textsuperscript{\orcidlink{0000-0002-9805-0078}},$^{3}$\newauthor
Johan De Keyser\,\textsuperscript{\orcidlink{0000-0003-4805-5695}},$^{4}$
Stephen A. Fuselier\,\textsuperscript{\orcidlink{0000-0003-4101-7901}},$^{5,6}$
Tamas I. Gombosi\,\textsuperscript{\orcidlink{0000-0001-9360-4951}},$^{3}$
Murthy S. Gudipati\,\textsuperscript{\orcidlink{0000-0001-5992-373X}},$^{7}$\newauthor
Nora Hänni\,\textsuperscript{\orcidlink{0000-0002-8253-6436}},$^{1}$
Kristina A. Kipfer\,\textsuperscript{\orcidlink{0000-0002-6544-9824}},$^{1,8}$
Niels F. W. Ligterink\,\textsuperscript{\orcidlink{0000-0002-8385-9149}},$^{1}$
Daniel R. Müller\,\textsuperscript{\orcidlink{0000-0003-3704-6865}},$^{1}$\newauthor
Yinsi Shou\,\textsuperscript{\orcidlink{0000-0002-5765-9231}},$^{3}$
and Susanne F. Wampfler\,\textsuperscript{\orcidlink{0000-0002-3151-7657}}$^{9}$
\\
\\
$^{1}$Space Research \& Planetary Sciences, Physics Institute, University of Bern, Sidlerstrasse 5, CH-3012 Bern, Switzerland\\
$^{2}$Laboratoire Atmosphères, Milieux, Observations Spatiales, Institut Pierre Simon Laplace, CNRS, Université Pierre et Marie Curie, 4 Avenue de Neptune,\\ F-94100 Saint-Maur, France\\
$^{3}$Department of Climate and Space Sciences and Engineering, University of Michigan, 2455 Hayward, Ann Arbor, MI 48109, USA\\
$^{4}$Royal Belgian Institute for Space Aeronomy, BIRA-IASB, Ringlaan 3, B-1180 Brussels, Belgium\\
$^{5}$Space Science Directorate, Southwest Research Institute, 6220 Culebra Rd., San Antonio, TX 78228, USA\\
$^{6}$Department of Physics and Astronomy, The University of Texas at San Antonio, San Antonio, TX 78249, USA\\
$^{7}$Science Division, Jet Propulsion Laboratory, California Institute of Technology, 4800 Oak Grove Drive, Pasadena, CA 91109, USA\\
$^{8}$NCCR PlanetS, Gesellschaftsstrasse 6, 3012 Bern, Switzerland\\
$^{9}$Center for Space and Habitability, University of Bern, Gesellschaftsstrasse 6, CH-3012 Bern, Switzerland
}

\date{Accepted XXX. Received YYY; in original form ZZZ}

\pubyear{2023}

\begin{document}
\label{firstpage}
\pagerange{\pageref{firstpage}--\pageref{lastpage}}
\maketitle

\begin{abstract}
ESA's Rosetta spacecraft at comet 67P/Churyumov-Gerasimenko (67P) was the first mission that accompanied a comet over a substantial fraction of its orbit. On board was the ROSINA mass spectrometer suite to measure the local densities of the volatile species sublimating from the ices inside the comet's nucleus. Understanding the nature of these ices was a key goal of Rosetta. We analyzed the primary cometary molecules at 67P, namely H$_2$O and CO$_2$, together with a suite of minor species for almost the entire mission. Our investigation reveals that the local abundances of highly volatile species, such as CH$_4$ and CO, are reproduced by a linear combination of both H$_2$O and CO$_2$ densities. These findings bear similarities to laboratory-based temperature programmed desorption experiments of amorphous ices and imply that highly volatile species are trapped in H$_2$O and CO$_2$ ices. Our results do not show the presence of ices dominated by these highly volatile molecules. Most likely, they were lost due to thermal processing of 67P's interior prior to its deflection to the inner solar system. Deviations in the proportions co-released with H$_2$O and CO$_2$ can only be observed before the inbound equinox, when the comet was still far from the sun and the abundance of highly volatile molecules associated with CO$_2$ outgassing were lower. The corresponding CO$_2$ is likely seasonal frost, which sublimated and lost its trapped highly volatile species before re-freezing during the previous apparition. CO, on the other hand, was elevated during the same time and requires further investigation.
\end{abstract}



\section{Introduction}\label{sec:Introduction}

Comets are remnants of the formation of our solar system \citep{Weissman2020}. As they spend their lifetime predominantly far away from the sun, they belong to the most pristine objects in the solar system. Due to this, observing comets reveals key information about the initial composition of the protoplanetary disk. To date, numerous comets have been observed remotely or have even been visited by spacecraft. The extensive body of cometary coma observations revealed a plethora of volatile species which originate from sublimation of ices inside their nuclei, on their surfaces, or from icy grains in the dust coma \citep[see, e.g.,][]{AHearn2011}. The most abundant molecules are water (H$_2$O), carbon dioxide (CO$_2$), and carbon monxide (CO), cf. \cite{AHearn2012} and \cite{Biver2019a}. Also a suite of minor volatiles have been observed, these include methane (CH$_4$), ethane (C$_2$H$_6$), propane (C$_3$H$_8$), molecular oxygen (O$_2$), and methanol (CH$_3$OH) among many others \citep{Bockelee2004,DelloRusso2016a,Bieler2015a,Schuhmann2019,Rubin2020}. All these volatile species cover a wide range in sublimation temperatures, from about $\SI{20}{\kelvin}$ to the sublimation of water at around $\SI{140}{\kelvin}$ \citep{Fray2009}.

Given that the vast majority of the relative abundances of these species have been derived from measurements of the gaseous coma surrounding the nucleus and mostly during the most active phase near perihelion, important questions remain unanswered. For instance, how are these volatiles stored in cometary ices inside the nucleus? There are several concurrent theories and the debate is still ongoing. Cometary ices may have formed through freeze-out in the protosolar nebula (PSN) with possible trapping of minor species in clathrates \citep{Luspay2016}. The other possibility is the inheritance of amorphous cometary ices from stages prior to the formation of the solar system, e.g, from the prestellar core stage or interstellar medium (ISM), before incorporation into the nucleus \citep{Altwegg2017a}. Both scenarios have been investigated extensively using numerical models and, where possible, with laboratory measurements \citep{Laufer2017,BarNun2007,Mousis2018}. Still, the debate remains to be settled. In particular because direct observations of pristine cometary ices are very limited and most evidence is derived from measurements of the gases in the coma.

Between 2014 -- 2016, the European Space Agency's Rosetta mission followed comet 67P/Churyumov-Gerasimenko (hereafter 67P) for over 2~years and carried out a close inspection of its nucleus and surrounding gas and dust coma \citep{Taylor2017}. Rosetta was designed to tackle a whole set of science goals, among them the determination of the composition of volatiles and the investigation of outgassing activity and associated seasonal effects \citep{Glassmeier2007}. Rosetta provided numerous new and surprising insights \citep{Fulle2016}, nevertheless, a number of questions remain in addition to new ones raised, in particular regarding activity and the origin and processing of the ices inside the comet's nucleus \citep{Thomas2019}. 

Understanding cometary activity requires detailed knowledge of the composition of the outgassing layer near the surface of the nucleus. For this purpose we analyze data from the Rosetta/ROSINA mass spectrometer suite \citep[Rosetta Orbiter Spectrometer for Ion and Neutral Analysis;][]{Balsiger2007} and then compare our results to the scenario of volatiles trapped in amorphous ice. In the following, we first introduce the Rosetta mission to comet 67P in section~\ref{sec:Rosetta}. Afterwards, we review some of the findings from laboratory experiments in section~\ref{sec:TPD}. The measurements obtained by Rosetta follow in section~\ref{sec:Observations} and a discussion with regards to the trapping of volatiles follows in section~\ref{sec:Discussion}. A summary of the major findings in section~\ref{sec:Conclusions} concludes the paper.

\section{Rosetta mission}\label{sec:Rosetta}
The Rosetta mission was launched on 2~March 2004 from Kourou, French Guayana, and arrived at comet 67P in early August 2014. For the next 25~months, Rosetta followed the comet along its orbit around the sun and carried out an in-depth investigation of its nucleus and surrounding gas, dust, and plasma environment \citep{Taylor2017}. During that time, 67P covered a heliocentric distance from well beyond $\SI{3.5}{\astronomicalunit}$ inbound, through perihelion at $\SI{1.24}{\astronomicalunit}$ in August 2015, and outbound to almost $\SI{4}{\astronomicalunit}$ again. Rosetta remained within a few several tens to hundreds of kilometers from the nucleus for the majority of the mission, depending on the outgassing activity of the comet. A substantial fraction was spent in gravitationally bound orbits within $\SI{30}{\kilo\meter}$ from the nucleus and in the terminator plane. Combined with the comet's rotation, Rosetta covered the entire surface and repetitively passed over the same sub-spacecraft latitudes and longitudes which allowed to study the long-term evolution of the nucleus and its surroundings. The mission was terminated by the end of September 2016, when the Rosetta spacecraft itself landed on the surface of the comet. 

Among the payload instruments on the orbiter was ROSINA \citep{Balsiger2007} which was dedicated to the in situ measurement of the major and minor volatile species at the location of the spacecraft \citep{Altwegg2019}. In the following, key aspects of the ROSINA instrument suite (subsection~\ref{subsec:ROSINA}) and the target, comet 67P (subsection~\ref{subsec:67P}) are introduced.

\subsection{The ROSINA instrument}\label{subsec:ROSINA}
The Double Focussing Mass Spectrometer (DFMS) and the COmet Pressure Sensor (COPS) were both part of the ROSINA instrument suite \citep{Balsiger2007}. DFMS was a mass spectrometer used to obtain the relative abundances of the volatile species at the location of Rosetta, in the coma of the comet. DFMS measured sequentially around each integer mass-per-charge ratio. Hence, the different volatile species in the coma were not acquired at the same time, e.g., there was a time difference of $\SI{13}{\minute}$ between the major coma species H$_2$O and CO$_2$. Therefore, the measured signals of a given species were linearly interpolated in time and normalized to the water signal to obtain relative abundances. Afterwards, the absolute densities were derived by scaling to the total density measured by COPS while maintaining the above-discussed relative abundances \citep{Rubin2019a}. 

Both DFMS and COPS had large fields-of-view (FoV), which most of the time, due to the cometocentric distance of Rosetta, covered much more than the entire nucleus of 67P. Thus it is not possible to pinpoint the exact origin of the molecules measured by ROSINA on the nucleus itself, but at least an approximate location can be identified \citep{Lauter2020,Combi2020}. In this work, however, we focus on the analysis of the local gas measurements at Rosetta instead of the source distribution on the nucleus.

More details about ROSINA DFMS and COPS and the data treatment can be found in section~\ref{app:ROSINA} in the appendix.

\subsection{Comet 67P}\label{subsec:67P}
Comet 67P, a Jupiter-family Comet (JFC) on its current orbit since a close encounter with Jupiter in 1959 \citep{Maquet2015}, has an orbital period of 6.4~years and a perihelion distance of $\SI{\sim1.24}{\astronomicalunit}$. The comet has a pronounced bi-lobate shape \citep{Sierks2015} and a rotation period that dropped from 12.4 to 12.0~hours during the inbound phase due to torques induced by outgassing, mostly during the peak activity period around perihelion \citep{Kramer2019}. Comet 67P's highly irregular shape most likely resulted from a collisional merger of two cometesimals \citep{Jutzi2015,Massironi2015}. The physical dimensions are approximately $\SI{4}{\kilo\meter}$ along the long axis and roughly 2 to $\SI{3}{\kilo\meter}$  in perpendicular direction. This, combined with the obliquity of 67P's rotation axis of $\SI{52}{\degree}$ \citep{Sierks2015}, leads to pronounced seasonal outgassing \citep{Hassig2015}. Summer on the northern hemisphere lasts for more than 5~years and covers the portion of the orbit away from the sun. The southern hemisphere, on the other hand, exhibits a short but intense summer lasting less than 1~year and includes perihelion. As a result, most of the gas and dust activity and hence also erosion occurs on the southern hemisphere \citep{Keller2015a,Keller2017}.

A number of neutral gas species' local and column densities have been monitored throughout the Rosetta mission by a suite of instruments. This includes Rosetta/MIRO \citep[Microwave Instrument on the Rosetta Orbiter;][]{Marshall2017,Biver2019b}, Rosetta/VIRTIS \citep[Visual IR Thermal Imaging Spectrometer;][]{Bockelee2015b, Bockelee2016}, and Rosetta/ROSINA \citep{Rubin2019b, Lauter2020} as well as multi-instrument studies including Rosetta/RPC, the Japanese PROCYON/LAICA wide-field imager, and ground-based observations of the dust activity \citep{Hansen2016, Combi2020}.

The major volatile species in 67P's coma are H$_2$O and CO$_2$. Additionally, CO, O$_2$, and H$_2$S (hydrogen sulfide) are present on the level of up to a few percent by number with respect to water \citep{Lauter2020,Combi2020}. All the other volatile species' abundances are below the percent level \citep{Rubin2019a}. No apparent difference in composition has been reported for the two lobes. Also the deuterium-to-hydrogen ratio in water, found to be elevated in comet 67P \citep{Altwegg2015}, was homogeneous within errors between both lobes \citep{Schroeder2019b}. Strong compositional heterogeneity in the gas coma, however, was observed above the northern versus the southern hemisphere due to the aforementioned obliquity of the rotation axis and associated seasons and is hence most likely of evolutionary origin \citep{LeRoy2015} .

The complex shape and orientation of the spin axis results in not only a complicated seasonal but also diurnal outgassing cycle with strongly varying relative abundances of coma volatiles \citep{Hassig2015,LeRoy2015}. Depending on the location of the Rosetta spacecraft, combined with the orientation of the nucleus, the main volatile in the coma could either be H$_2$O or CO$_2$. Based on early mission data, \cite{LeRoy2015} showed that the relative gas abundances in the coma with respect to water changed substantially from the, at the time, illuminated northern summer hemisphere to the southern winter hemisphere. For instance, the CO$_2$/H$_2$O ratio changed by a factor $\sim$30, the CO/H$_2$O ratio by $\sim$7, the CH$_4$/H$_2$O ratio by $\sim$4, and the C$_2$H$_6$/H$_2$O ratio by $\sim$10. \cite{Hassig2015} suggested that these variations are governed by temperature variations, which depend on the depth of the outgassing layer and the changing illumination of the rotating nucleus.

\cite{Luspay2015, Luspay2019} studied correlations of different volatiles above the southern hemisphere in- and outbound beyond $\SI{3}{\astronomicalunit}$. The analysis included HCN (hydrogen cyanide), CH$_3$OH, C$_2$H$_6$, CH$_4$ and CO in comparison to a mix of the primary volatiles H$_2$O and CO$_2$. In particular, CH$_4$ exhibited a distinct diurnal outgassing pattern which did no follow either of the major species. The authors could then attribute the pronounced changes in the abundances of these volatiles from pre- to post-equinox to seasonally driven heterogeneity of the nucleus caused by strong erosion of the southern hemisphere.

\cite{Gasc2017} investigated a suite of 8 species (H$_2$O, O$_2$, NH$_3$ (ammonia), H$_2$S, CH$_4$, HCN, CO$_2$, and CO) and showed that their outgassing pattern did not correlate with the sublimation temperature or polarity of the molecules. During the outbound journey, the different species' outgassing activity showed a dependence on the heliocentric distance between H$_2$O (r$_\textrm{h}^{-7.32\pm0.04}$) and CO$_2$ (r$_\textrm{h}^{-2.18\pm0.04}$). The authors interpreted this result to be the consequence of two different ice phases, H$_2$O and CO$_2$, with all the other species trapped in different relative abundances in these ices, consistent with the re-analysis of O$_2$ by \cite{Luspay2022}.

67P also exhibits a hemispherical dust transport across the surface due to inhomogeneous outgassing and dust activity \citep{Thomas2015,Rubin2014}. Such transport of material has also been observed at other comets, such as the redeposition of icy grains at comet 103P/Hartley~2 \citep{AHearn2011}. At 67P and in particular during the short but intense summer period around perihelion, substantial erosion occurs on the southern hemisphere \citep{Keller2015a}. Dust is lifted, transported and redeposited on the northern hemisphere \citep{Keller2017}.

\section{Temperature-programmed desorption experiments}\label{sec:TPD}

Temperature-programmed desorption (TPD) measurements are an established laboratory technique to study the desorption processes of ices in the astrophysical context including comets \citep{Burke2010,Minissale2022}. A number of relevant species have been co-deposited together with H$_2$O \citep[e.g.,][]{Collings2004,Martin2014,Almayrac2022} and CO$_2$ \citep[e.g.,][]{Simon2019} at different trapping temperatures. The release of volatiles was then studied during a defined, most often linear, warm-up phase, hence the name of the technique. Key input parameters for such experiments include the gas composition, the temperature of the cold substrate on which the gas was frozen out, the thickness of the accumulated ice, and the heating rate used during the subsequently monitored desorption process \citep{Collings2004}. 

In the programmed warm-up phase, species are released in different temperature regimes. For highly volatile molecules, like CH$_4$, CO, N$_2$ (molecular nitrogen), and O$_2$, a first part is released at or near its characteristic sublimation temperature \citep[see, e.g,][]{Fray2009}. At elevated temperatures, the portion trapped in other species then co-desorbs at or near the sublimation temperature of the corresponding host species, e.g., CO$_2$ and H$_2$O. In the following we will refer to species with a pure ice sublimation temperature below that of CO$_2$ as highly volatile species.

Of particular interest for our work are the two most abundant species in comet 67P because they make up for the matrix/host phases: H$_2$O and CO$_2$ \citep{Rubin2019a,Combi2020,Lauter2020}. In the case of H$_2$O, the situation can be quite complex: trapped species are also released during the phase transition from low porosity amorphous to crystalline water ice, at lower temperatures than the main sublimation of water \citep{Burke2010}. The release during the crystallization of amorphous water ice is also known as molecular volcano desorption \citep{Smith1997}. Additionally, whether a species is released entirely during the water ice phase transition or co-desorbs with H$_2$O also depends on the thickness of the ice: \cite{Notesco2005} found that thin water ice films ($\SI{0.1}{\micro\meter}$) lose most trapped impurities while thick water ices  ($\SI{5}{\micro\meter}$) may quench the outward diffusion and the gas release then follows together with the main water ice sublimation. However, low levels of water desorption can occur before the amorphous to crystalline phase transition of water \citep{Gudipati2023}. This is of particular importance when ample time is available, as in the case of the ISM. In this situation the bulk of the H$_2$O outgassing occurs already at much lower temperatures. The same applies to CO$_2$ and other volatiles \citep{Ligterink2023}.

\begin{center}
\begin{figure}
\includegraphics[width=1.0\columnwidth]{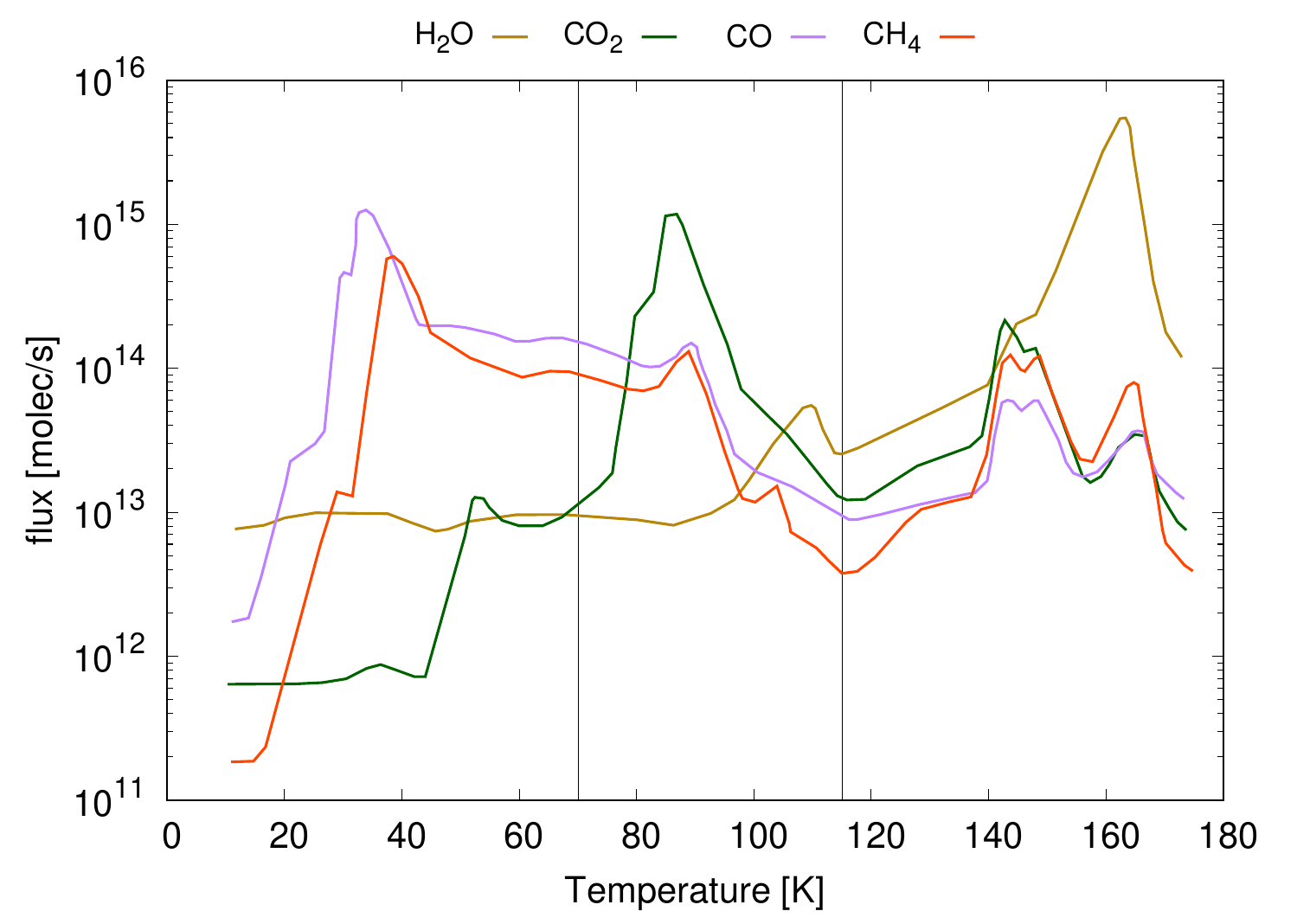}
\caption{TPD measurement of H$_2$O, CO$_2$, CO, and CH$_4$ after deposition at $\SI{10}{\kelvin}$ in the ratio 65:10:15:10, respectively. Data obtained from \protect\cite{Kouchi1995}. The two vertical lines separate the three temperature regimes introduced in Table~\ref{tab:TPD_ranges}. The portions released in the three temperature ranges are listed in Table~\ref{tab:TPD_Kouchi}.}
\label{fig:TPD_Kouchi}
\end{figure}
\end{center}

TPD experiments are crucial for the case of comets. For instance, \cite{Kouchi1990} investigated a mixture consisting of H$_2$O, CO$_2$, CO, and CH$_4$ deposited at $\SI{10}{\kelvin}$ in the ratio 65:10:15:10.  All four species are commonly observed in comets \citep{AHearn2012,DelloRusso2016a}.The results are reproduced in Fig.~\ref{fig:TPD_Kouchi} and we will refer to this experiment later on.

\begin{center}
\begin{figure}
\includegraphics[width=1.0\columnwidth]{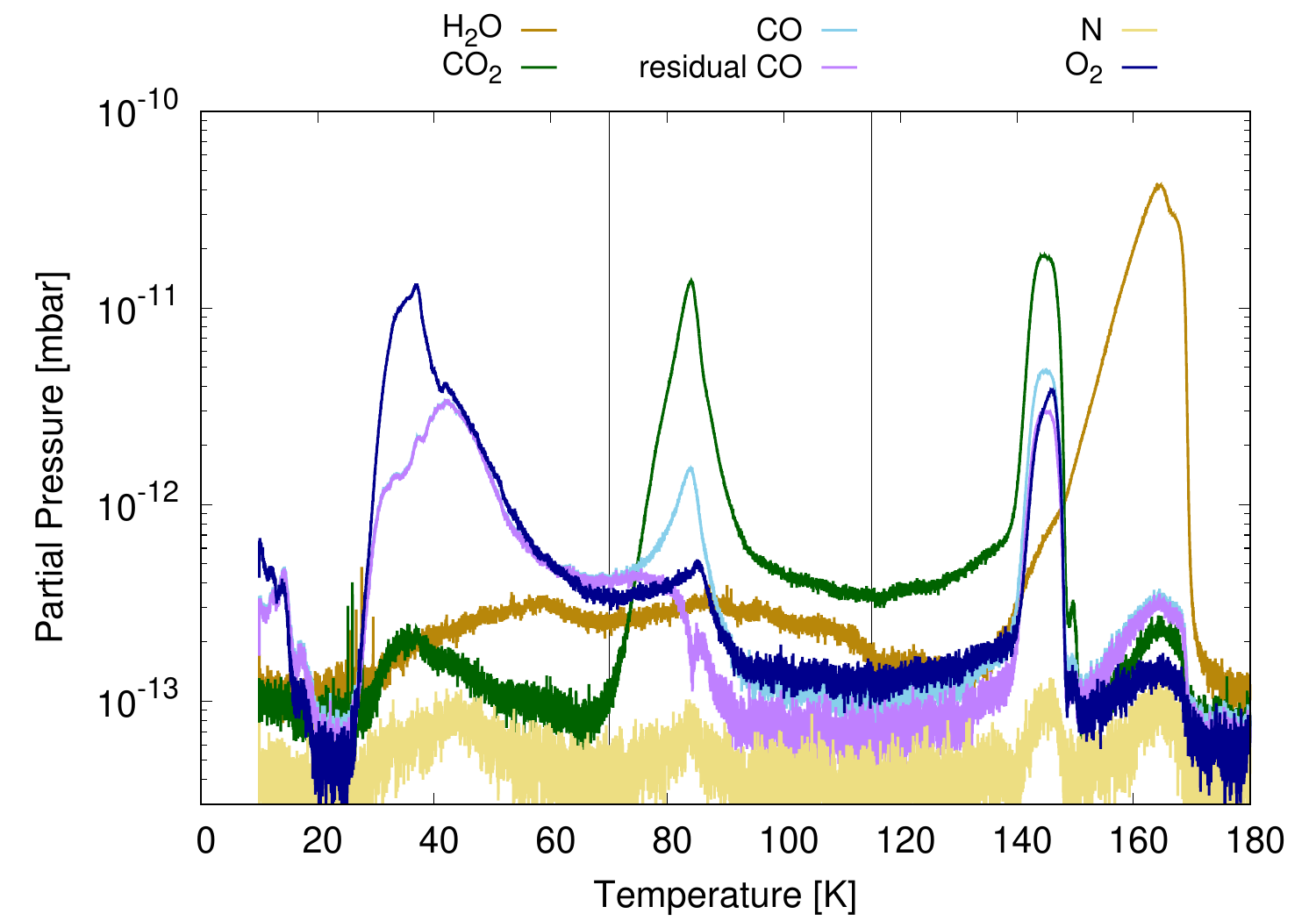}
\caption{TPD measurement of H$_2$O, CO$_2$, CO, and O$_2$ in the ratio 60:20:10:10, respectively, with a heating rate of $\SI{0.5}{\kelvin\per\minute}$ \citep{Gudipati2023}. For CO also the residual signal has been derived, i.e., after subtraction of the contribution by fragmentation of $\mathrm{CO}_2$. The experiment also tracked the chamber background species N, a fragment of N$_2$. The two vertical lines separate the three temperature regimes introduced in Table~\ref{tab:TPD_ranges}. The portions released in the three temperature ranges are listed in Table~\ref{tab:TPD_Gudipati}.}
\label{fig:TPD_Gudipati}
\end{figure}
\end{center}

Another relevant experiment has recently been conducted by \cite{Gudipati2023} and is reproduced in Fig.~\ref{fig:TPD_Gudipati}. It included a mixture of H$_2$, CO$_2$, CO, and O$_2$ in the ratio 60:20:10:10. The thickness was estimated to be about $\SI{0.25}{\micro\meter}$ (approximately 1000 monolayers) based on calibrated reference water ice samples. In particular, CO and O$_2$ seem to be much better trapped in H$_2$O as opposed to CO$_2$, i.e., they mostly outgass from their CO- and O$_2$-dominated ice phases below $\SI{70}{\kelvin}$ and then again above $\SI{140}{\kelvin}$. Only a small fraction is associated with CO$_2$. Whether or not these ratios change as a function of ice thickness remains to be investigated. The portion of the outgassing occurring from the O$_2$-dominated ice, however, depends on the initial trapping temperature. In their experiment $\SI{10}{\kelvin}$ was used. Furthermore, also residual species from the vacuum chamber background were tracked. This included N, a fragment of N$_2$ through dissociative electron impact ionization. As in the experiments by \cite{Kouchi1995}, a cryostat set to $\SI{10}{\kelvin}$ was used and also the relative amounts of H$_2$O, CO$_2$, and CO were similar. 

Additional molecules that have been studied in mixtures of H$_2$O and CO$_2$, are NH$_3$ and CH$_3$OH \citep{Martin2014}. At comet 67P, both molecules can be considered trace species in the ices of the outgassing layer with abundances below the percent level with respect to H$_2$O \citep{Rubin2019a,Lauter2020}.

\cite{Kouchi2001} have shown that impurities >2\% in the H$_2$O ices will turn the exothermic behavior of the amorphous to crystalline ice transitions endothermic, unlike the situation for pure water ice \citep[see, e.g.,][]{Gudipati2023}. Therefore, a runaway phase transition of amorphous to crystalline water ice cannot be expected for an ice composition characteristic for comets and is hence not a major driver of the outgassing activity. 

Based on the experiments by \cite{Kouchi1990} and  \cite{Gudipati2023}, which are also qualitatively in agreement with \cite{Martin2014}, we define the three temperature ranges listed in Table~\ref{tab:TPD_ranges} (cf. Figs.~\ref{fig:TPD_Kouchi} and \ref{fig:TPD_Gudipati}). The selection of temperature limits is somewhat arbitrary, but for now we will stick to this simplistic approach as it allows for the qualitative comparison of the different datasets. For the following discussion, however, it is important to note that the numbers may change slightly in case different limits are selected. The general behavior, however, remains the same.

\begin{table}
\caption{Outgassing behavior of the H$_2$O, CO$_2$, CO, and CH$_4$ mixture (ratio 65:10:15:10) and the H$_2$O, CO$_2$, CO, and O$_2$ mixture (ratio 6:2:1:1 with residual N$_2$) according to the TPD measurements shown in Figs.~\ref{fig:TPD_Kouchi} and \ref{fig:TPD_Gudipati} and split into three temperature ranges, separated by the species governing the outgassing: highly volatiles, CO$_2$, or H$_2$O.}
\begin{tabularx}{\columnwidth}{lX}
 <~$\SI{70}{\kelvin}$ & Ices dominated by highly volatile species, CH$_4$, N$_2$, CO, and O$_2$, are being released. Marginal co-release of the the lesser volatile molecules CO$_2$ and H$_2$O. \\ 
  70--$\SI{115}{\kelvin}$ & CO$_2$ outgassing with co-desorption of the trapped CO, N$_2$, CH$_4$, and small amounts of O$_2$. Marginal co-release of the lesser volatile H$_2$O. \\ 
  >~$\SI{115}{\kelvin}$ & CH$_4$, CO, O$_2$, N$_2$, and CO$_2$ are being released during the amorphous to crystalline H$_2$O ice transition, i.e., the molecular volcano at about $\SI{145}{\kelvin}$, with a minor contribution of H$_2$O. The bulk sublimation of H$_2$O then occurs a few degrees above that with additional release of small amounts of highly volatile molecules. \\ 
\end{tabularx}
\label{tab:TPD_ranges}
\end{table}

Table~\ref{tab:TPD_Kouchi} lists what fraction of the volatiles shown in Fig.~\ref{fig:TPD_Kouchi} is released in each temperature interval from Table~\ref{tab:TPD_ranges}. For instance, almost 70\% of the CH$_4$ was in its own phase and then released at or near its pure ice sublimation temperature \citep[cf.,][]{Fray2009}, about 14\% was released together with CO$_2$, and the rest was associated with H$_2$O sublimation, either in the molecular volcano or co-release \citep{Kouchi1995}. A very similar picture arises for CO. Furthermore, half of the CO$_2$ was released together with H$_2$O in these experiments. Similarly, the results from Fig.~\ref{fig:TPD_Gudipati} are collected in Table~\ref{tab:TPD_Gudipati}. 

Both measurements were obtained using the same deposition temperature and similar relative amounts of host species CO$_2$  and H$_2$O. The results generally agree but there are also some differences, for instance, the fraction of the CO$_2$ co-desorbing with H$_2$O (in the temperature regime $\SI{>115}{\kelvin}$) is larger in the TPD measurements by \cite{Gudipati2023}. On the other hand, the CO fraction released together with CO$_2$ (i.e., in the temperature regime 70-$\SI{115}{\kelvin}$) is smaller than observed by \cite{Kouchi1995}, especially after subtraction of the fragmentation contribution of $\mathrm{CO}_2$ through dissociative electron impact ionization. Both cases in Figs.~\ref{fig:TPD_Kouchi} and \ref{fig:TPD_Gudipati} show a continuum-type outgassing of CO, O$_2$, and CH$_4$ extending from the $\SI{<70}{\kelvin}$ regime well into the CO$_2$-dominated regime. In parts this has to do with the heating rate used for the experiments, which is probably faster compared to the situation at a comet (see later discussion in section~\ref{sec:Observations}). Furthermore, \cite{Gudipati2023} also referred to a limited pumping efficiency of the CO in their vacuum system. In fact, at pressures below $\SI{1e-8}{\milli\bar}$ during the warm-up phase, it is quite impossible to remove completely the gaseous highly volatile molecules on the time scale of the experiment. Therefore, surface-bound CO may linger at elevated temperatures after its release, resulting in a higher mass spectrometric signal. This is a commonly known limitation of the TPD technique. Nevertheless, we did not modify our results for such an effect, but taking all this into account, the relative fraction of CO, O$_2$, and CH$_4$ that is co-released with CO$_2$ is likely lower than the numbers provided in Tabs.~\ref{tab:TPD_Kouchi} and \ref{tab:TPD_Gudipati}. These are complications that we have to keep in mind when putting our measurements at 67P in the context of the laboratory experiments.

\begin{table}
\begin{center}
\caption{Normalized fractions of release of H$_2$O, CO$_2$, CO, and CH$_4$ in the indicated temperature ranges during the TPD experiment by \protect\cite{Kouchi1995} shown in Fig.~\ref{fig:TPD_Kouchi}.}
\begin{tabular}{ r c c c c}
Temperature range
& H$_2$O
& CO$_2$
& CO
& CH$_4$ \\ \hline

<~$\SI{70}{\kelvin}$
& 0.012
& 0.017
& 0.756
& 0.612 \\

70 -- $\SI{115}{\kelvin}$
& 0.018
& 0.782
& 0.172
& 0.203 \\

>~$\SI{115}{\kelvin}$
& 0.970
& 0.201
& 0.071
& 0.185 \\

\end{tabular}
\label{tab:TPD_Kouchi}
\end{center}
\end{table}

\begin{table}
\begin{center}
\caption{Normalized fractions of release of H$_2$O, CO$_2$, CO, N$_2$ (based on fragment N), and O$_2$ in the indicated temperature ranges during the TPD experiment from \protect\cite{Gudipati2023} shown in Fig.~\ref{fig:TPD_Gudipati} and after offset-subtraction. For CO the residual signal has been used, i.e., after subtraction of the contribution by fragmentation of $\mathrm{CO}_2$.}
\begin{tabular}{ r c c c c c}
Temperature range
& H$_2$O
& CO$_2$
& CO
& N$_2$ (N)
& O$_2$ \\ \hline

<~$\SI{70}{\kelvin}$
& 0.017
& 0.015
& 0.700
& 0.388
& 0.816 \\

70 -- $\SI{115}{\kelvin}$
& 0.020
& 0.435
& 0.071
& 0.191
& 0.056 \\

>~$\SI{115}{\kelvin}$
& 0.963
& 0.550
& 0.230
& 0.421
& 0.128 \\

\end{tabular}
\label{tab:TPD_Gudipati}
\end{center}
\end{table}

\section{Observations at comet 67P}\label{sec:Observations}
ROSINA detected and monitored numerous volatiles throughout the Rosetta mission. For the first time, changes in the composition or ratios of the outgassing can be studied with high temporal resolution and excellent coverage in time. In the following, the acquired dataset is used to study the behavior of the different volatiles with respect to each other. But first, we will summarize a suite of observations that are relevant for our study.

\subsection{Supporting observations}\label{subsec:SupportingObs}
For comets, we only have limited information about the temperature of the outgassing layer. \cite{Gulkis2015} presented millimeter (mm) and submillimeter (smm) brightness temperatures of the northern and the southern hemispheres in the range of 60-$\SI{180}{\kelvin}$, measured by MIRO with spatial resolution down to $\SI{40}{\meter}$ when 67P was still beyond $\SI{3.2}{\astronomicalunit}$. During that time, the northern hemisphere was in summer and, hence, elevated temperatures were measured compared to the south (cf. section~\ref{subsec:67P}). The brightness temperature showed strong spatial variations, mostly associated with illumination and self-shadowing due to the comet's complex shape. On a longer time scale, this is furthermore coupled with a varying dust cover, for instance through transport and redeposition, changing sub-solar latitude, and variation in solar irradiation as a function of the changing heliocentric distance along the comet's orbit. 

The MIRO mm temperatures are lower than the smm temperatures, hence mark a decrease in temperature with depth from the comet's surface. In comparison, the maximum diurnal surface temperature in the August to September 2014 time frame was on the order of $\SI{230}{\kelvin}$ inferred from VIRTIS measurements in the infrared spectral wavelength range of 4.5 -- $\SI{5.1}{\micro\meter}$ \citep{Tosi2019}. Compared to the MIRO observations, VIRTIS temperatures would be much closer to the surface, i.e., a few tens of microns versus approximately a centimeter, respectively. In combination, these results are consistent with a very porous, low bulk thermal inertia nucleus which indicates that pristine material may still be present at relatively shallow depth \citep{Groussin2019}. While the temperatures derived by MIRO may not reflect the actual outgassing layers, \cite{Gulkis2015} suggested that the smm and mm radiation is affected by the temperature at the diurnal thermal skin depth, which they estimated to be on the order of 1 to $\SI{2}{\centi\meter}$ for 67P at $\SI{3.84}{\astronomicalunit}$. Other components affecting the temperature are the seasonal skin depth ($\SI{\sim1}{\meter}$) and the isothermal layer underneath.

Closer to perihelion ($\SI{1.24}{\astronomicalunit}$), however, the solar irradiation was substantially higher. Still, there may be regions on the comet that never reached the amorphous to crystalline ice transition temperature of water, or only during a limited time and within a limited heliocentric distance. This may still have resulted in a slow sublimation of the amorphous water ice together with all the trapped species within. An example is comet Hale-Bopp, for which water activity has been observed even beyond $\SI{4}{\astronomicalunit}$ \citep{Biver2002a}. On a temporal scale, thermal gradients up to $\SI{0.1}{\kelvin\per\minute}$ have been reported for 67P when looking at the averaged brightness temperature \citep{Gulkis2015}. Even higher thermal gradients, locally up to 2--$\SI{9}{\kelvin\per\minute}$, have been observed by VIRTIS on the surface of regions which experience sudden daytime shadowing \citep{Tosi2019}. Such thermal gradients may lead to strong thermal stresses, for instance in 67P’s neck area \citep[Hapi; cf. ][]{El-Maarry2015}, where also a water frost cycle has been observed \citep{DeSanctis2015}.

The gas measured by ROSINA may originate from all over the nucleus' surface facing Rosetta due to the large FoV (section~\ref{subsec:ROSINA}), even if most of it originates from a much more confined location \citep{Kramer2017,Lauter2019,Lauter2020,Combi2020}. But even in small areas the illumination conditions, dust cover, etc. can vary substantially, which greatly broadens up the parameter space for relevant laboratory ice experiments. It is therefore clear that there is no single TPD experiment covering the whole range of conditions encountered at any given point in time. Complications arise from, e.g., the distribution of the gas sources across the comet's complex shape coupled with its rotation and orientation of the rotation axis, the variation in heliocentric distance, dust coverage, and thermal gradients in its interior. Hence, it is unlikely that ROSINA measurements can distinguish outgassing caused by the molecular volcano or co-release during the water ice sublimation due to the rather close temperature and close source regions on the comet. Accordingly, ROSINA data does not reveal the exact structure and stratification of the ices. Therefore, when talking about trapping, we not only include ices with incorporated impurities, but also layered ices, where the upper layer quenches the outward diffusion of more volatile species from below. This is based on the thick and thin ice experiments performed by \cite{Notesco2005} and discussed above.

Still, a number of conclusions can be drawn for a comet like 67P. In the following, and according to Table~\ref{tab:TPD_ranges}, we simply consider the 70--$\SI{115}{\kelvin}$ temperature regime to be CO$_2$-dominated and driven. The $\SI{>115}{\kelvin}$ temperature regime is assumed to be H$_2$O-outgassing driven, including both the molecular volcano desorption and co-desorption (section~\ref{sec:TPD}). We assume that the major species, CO$_2$ and H$_2$O govern the outgassing behavior, but will also look for signs, or the absence thereof, of separate ice phases dominated by highly volatile species. 

With regards to the observations at 67P, the species provided in Tabs.~\ref{tab:TPD_Kouchi} and \ref{tab:TPD_Gudipati} are all relevant. First, they contain the two primary cometary coma molecules, H$_2$O and CO$_2$. Then CO, CH$_4$, N$_2$ (monitored through fragment N), and O$_2$. The corresponding pure ice sublimation temperatures are listed in Table~\ref{tab:results}. There are many more species observed in comets, however, if their sublimation temperature is higher than CO$_2$, and possibly even H$_2$O, they may not be efficiently co-released. For instance, Fig.~\ref{fig:TPD_Kouchi} shows that CO is released with H$_2$O but not very much the other way round. This, of course, is emphasized due to higher H$_2$O abundance in the experiment. The same applies to other combinations, e.g., both CO and CH$_4$ are released together with CO$_2$ but, in relative numbers, only little of the total CO$_2$ comes off together with CO or CH$_4$, respectively. Another example is CH$_3$OH which does no co-desorb with CO \citep{Ligterink2018}. And then there are also numerous species of even lower volatility than water, sometimes even associated with dust \citep{Altwegg2017b,Hanni2022}.

\subsection{H$_2$O and CO$_2$}\label{subsec:H2OCO2}

\begin{center}
\begin{figure*}
\includegraphics[width=1.0\textwidth]{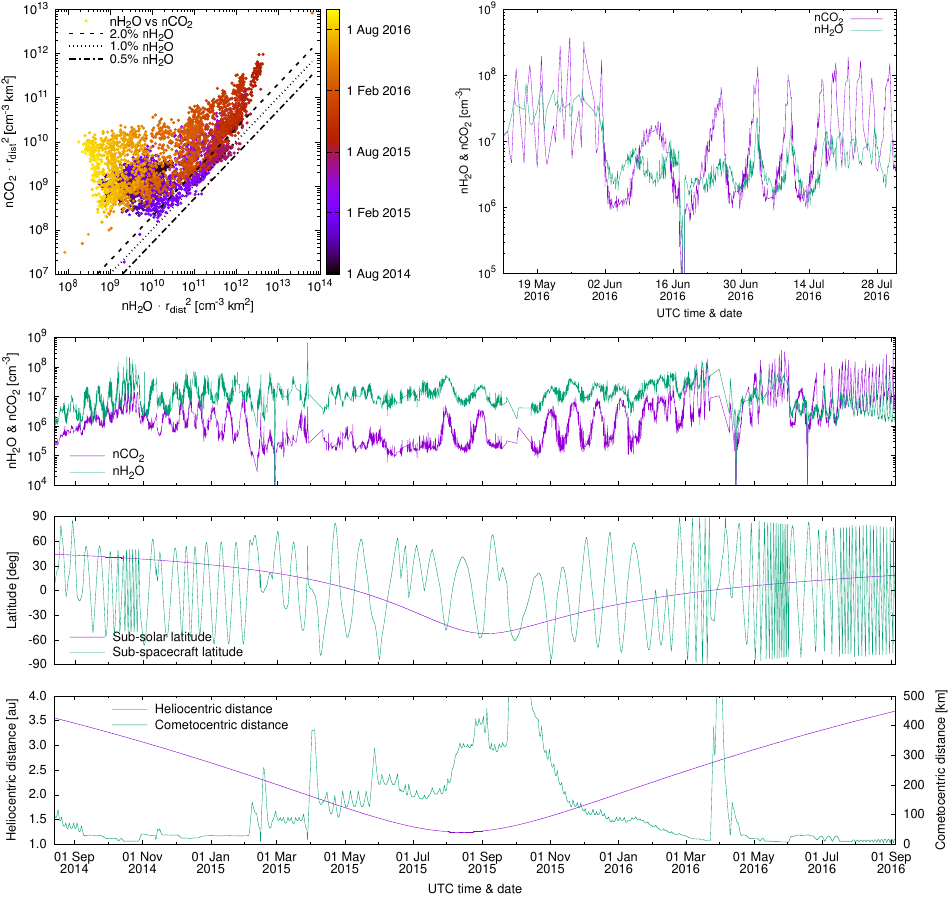}
\caption{Top left panel: Measured H$_2$O versus CO$_2$ densities between mid August 2014 (inbound at $\SI{3.6}{\astronomicalunit}$), though perihelion ($\SI{1.24}{\astronomicalunit}$), until September 2016 outbound at ($\SI{3.7}{\astronomicalunit}$) and multiplied by the cometocentric distance squared. The dashed and dotted lines mark the minimum CO$_2$ densities when 0.5\%, 1\%, or 2\% is (trapped in and) co-desorbing with H$_2$O. Top right panel: H$_2$O and CO$_2$ densities measured by ROSINA during close ellipses in summer 2016 when 67P was on its outbound journey near $\SI{3.2}{\astronomicalunit}$ from the sun. For readability the measurements have been connected by straight lines, despite the measurements being rather sparse during May/June 2016 as different measurement modes were executed, e.g., the search for noble gases \citep{Balsiger2015,Marty2017,Rubin2018}. The lower three panels, top to bottom: measured H$_2$O and CO$_2$ densities during the whole analyzed period as shown in the top left panel, the corresponding subsolar and sub-spacecraft latitudes, and the corresponding heliocentric distance of 67P and the distance between Rosetta and the comet used in the top left panel.}
\label{fig:H2OCO2}
\end{figure*}
\end{center}

The focus of this section is on the two main ice phases in 67P, H$_2$O and CO$_2$. The top left panel in Fig.~\ref{fig:H2OCO2} shows the densities of the two molecules measured by ROSINA at the location of Rosetta versus each other after multiplication by the squared distance to the comet, r$_\mathrm{dist}^2$, in the following called modified density. This modification has been applied to remove, to first order, the variation in cometocentric distance. The data starts mid August 2014, just after arrival at the comet and then extends to 5 September 2016, when a dust impact into the DFMS instrument hampered subsequent measurements by physically blocking the current of ionizing electrons \citep{Altwegg2017b}. As a result the data shown here covers a range of heliocentric distances, i.e., from $\SI{3.6}{\astronomicalunit}$ pre-perihelion to $\SI{3.7}{\astronomicalunit}$ post-perihelion. Otherwise, no limitation was applied regarding the position of the spacecraft with respect to the comet and hence the presented data covers a wide range of sub-spacecraft latitudes and longitudes as well as phase angles. 

The top right panel of Fig.~\ref{fig:H2OCO2} shows a time series of the H$_2$O and CO$_2$ densities measured by ROSINA during a period of several weeks late in the Rosetta mission. Both species show a very distinct behavior, especially between early and mid June 2016, a time when Rosetta moved from the terminator orbit (phase angle $\SI{90}{\degree}$) to a lower phase angle of about $\SI{44}{\degree}$ and then back again. Peaks in the measured densities occurred when Rosetta was passing above the continuously more active southern hemisphere past the outbound equinox (sub-solar latitude $\SI{>5}{\degree}$ north). Hence, CO$_2$ showed a stronger dependence on the sub-spacecraft latitude than H$_2$O. As a consequence, this period, which reveals a strong dichotomy between CO$_2$ and H$_2$O, is well suited to visualize the different behavior of the highly volatile species co-released with the two host species.

Fig.~\ref{fig:H2OCO2}  also shows the measured H$_2$O and CO$_2$ densities for the entire time period analyzed (including the detailed view in the top right panel). For most of the time throughout the Rosetta mission, H$_2$O was the dominant species and H$_2$O-outgassing followed more or less the sub solar latitude \citep{Combi2020,Lauter2020}. CO$_2$, on the other hand, was predominantly released from the southern hemisphere, in particular also during the early Rosetta mission when the southern hemisphere was only poorly illuminated \citep{Lauter2020}. The two bottom panels show the corresponding sub-solar and sub-spacecraft latitudes and the heliocentric and cometocentric distances.

As previously reported, e.g., by \cite{Hassig2015}, \cite{Gasc2017}, and \cite{Luspay2019} and in line with Fig.~\ref{fig:H2OCO2}, both primary cometary parent species, H$_2$O and CO$_2$, are only poorly correlated. Part of the remaining correlation shown in the left panel is caused by the strong variation in outgassing activity due to the change in heliocentric distance: both species have their peak outgassing coinciding about 2 weeks after perihelion (red color points in the upper right corner of the top left panel) whereas farther away (purple: pre-perihelion and yellow: post perihelion) the outgassing of both molecules was significantly lower \citep{Combi2020,Lauter2020}. 

This seasonal evolution is in agreement with observations by VIRTIS, which revealed a change in composition of the comet’s surface during the same time frame. \cite{Filacchione2020} and \cite{Ciarniello2022} showed that erosion of the nucleus, caused by enhanced gas and dust activity around perihelion, led to a blueing of the surface resulting from the exposure of more pristine icy layers. Far from the sun, dehydration of the surface combined with the redeposition of dust decreased the amount of ices in the surface layer.

Some degree of the correlation observed in the top left panel of Fig.~\ref{fig:H2OCO2} incurs from just the change in heliocentric distance and hence activity. Nevertheless, water and carbon dioxide abundances show strong variations with respect to each other, in fact the CO$_2$/H$_2$O ratio changes by more than two orders of magnitude. While early on in the Rosetta mission, CO$_2$ could be as low as a few percent with respect to H$_2$O, during the final months, when Rosetta was beyond $\SI{2.7}{\astronomicalunit}$ outbound, the carbon dioxide became the dominant molecule in the coma \citep{Lauter2020}. The top right panel, for instance, shows a period where local H$_2$O and CO$_2$ abundances were comparable, one dominating over the other at times and then vice versa.

CO$_2$ has always been present in the coma, even above the deposit layers in the northern hemisphere, where airfall is expected to be mostly depleted in CO$_2$ \citep{Davidsson2021a}. The top left panel of Fig.~\ref{fig:H2OCO2}  also shows lines representing lower limits for the CO$_2$ when assuming that 0.5\%, 1\%, or 2\% is (trapped in and) co-desorbing with H$_2$O and reveals no relative abundances of CO$_2$ with respect to H$_2$O below 0.5\%. 

As a consequence, from here on we assume that about 1\% of the carbon dioxide is trapped in water. We can compare the 1\% CO$_2$ (w.r.t. H$_2$O) to the total 7.5\% CO$_2$ from the integrated outgassing over the whole Rosetta mission \citep{Lauter2020}. This leaves 6.5\% CO$_2$ to be present in its separate (multilayer) phase, i.e., in a ratio of 1:6.5, see also Table~\ref{tab:results}. If, for the sake of comparison, we base our analysis on the bulk abundance of CO$_2$ of 4.7\% (w.r.t. H$_2$O), derived just before perihelion \citep{Rubin2019a}, the partitioning of CO$_2$ in H$_2$O versus a CO$_2$-dominated phase would be 1:3.7. TPD measurements show that CO$_2$ is easily trapped in H$_2$O. The experiment by \cite{Kouchi1995} in Table~\ref{tab:TPD_Kouchi} revealed a ratio of about 1:4 and \cite{Gudipati2023} in Table~\ref{tab:TPD_Gudipati} obtained approximately 1:1. These ratios depend on the deposition rate and temperature of the experiments in the laboratory and one should be careful in comparing them at face value.

Furthermore, there are also caveats associated with ROSINA measurements: H$_2$O and CO$_2$ were measured with a time difference of about $\SI{15}{\minute}$ and interpolated in time. Additionally, there are errors related to the measured ratios of $\sim20\%$ originating from sensitivity calibration of DFMS. Therefore, an even higher fraction of the carbon dioxide trapped in water cannot be excluded either.

In the next section, we will study the highly volatile molecules which are co-released with both the H$_2$O-dominated and the CO$_2$-dominated ice phases.

\subsection{Highly volatile molecules}\label{subsec:highvolatility}
Here we discuss the highly volatile molecules CO, CH$_4$, O$_2$, N$_2$, H$_2$S, C$_2$H$_6$, and C$_3$H$_8$ that were all commonly observed in the coma of 67P. For this analysis, we limited the dataset to the time interval when the southern hemisphere of the comet was more active than the northern hemisphere, i.e., starting from early February 2015 inside $\SI{2.4}{\astronomicalunit}$ inbound \citep{Lauter2020}. Only early on in the Rosetta mission, the northern hemisphere was the dominant source of volatiles in the coma. The southern hemisphere, on the other hand, showed the highest levels of activity and, hence, erosion due to the short but intense summer around perihelion \citep{Keller2015a}. As a consequence, the analyzed data mostly represents outgassing from fresh cometary ices that continuously became accessible. As a result the data shown here covers a range of heliocentric distances, i.e., from $\SI{2.4}{\astronomicalunit}$ pre-perihelion to $\SI{3.7}{\astronomicalunit}$ post-perihelion on 5 September 2016. Later on in the section~\ref{subsec:EarlyMission} we will then also discuss the omitted early period from August 2014 to February 2015.

\begin{table*}
\begin{center}
\caption{Relative abundances in 67P of H$_2$O, CO$_2$, a suite of species with volatility between H$_2$ and CO$_2$ (CH$_3$OH, HCN, NH$_3$), and highly volatile molecules (C$_2$H$_6$, C$_3$H$_8$, CH$_4$, O$_2$, CO), sorted by their pure ice sublimation temperature from \protect\cite{Fray2009} and \protect\cite{Behmard2019} for C$_3$H$_8$. The bulk abundances are taken from the mission integrated outgassing$^\mathrm{a)}$ from \protect\cite{Lauter2020} where available or from the relative outgassing period just before perihelion$^\mathrm{b)}$ from \protect\cite{Rubin2019a}. Columns four and five list the partitioning in the two main ice phases, H$_2$O and CO$_2$. The last column lists the maximum amounts co-released with H$_2$O (cf. Figs.~\ref{fig:H2OCO2} and \ref{fig:MaxTrapH2O}), e.g., up to 2\% of the total 7.5\% of the CO$_2$ can be trapped in or underneath water which corresponds to 27\% of the total CO$_2$.}
\begin{tabular}{ c  S[table-format=3.0] S[table-format=3.3]  S[table-format=3.1]   S[table-format=3.1]  S[table-format=3.1]  }
Species &
\multicolumn{1}{c}{Sublimation} &
\multicolumn{1}{c}{bulk abun-} &
\multicolumn{1}{c}{portion with} &
\multicolumn{1}{c}{portion with} &
\multicolumn{1}{c}{Max portion}  \\

&
\multicolumn{1}{c}{temperature [K]} &
\multicolumn{1}{c}{dance [\%]} &
\multicolumn{1}{c}{H$_2$O [\%]} &
\multicolumn{1}{c}{CO$_2$ [\%]} &
\multicolumn{1}{c}{with H$_2$O [\%]} \\ \hline

H$_2$O &
144 &
100  &
100  &
0  &
\text{--} \\

CH$_3$OH &
142 &
0.54$^\mathrm{a)}$ &
\text{--} &
\text{--} &
37 \\

HCN &
126 &
0.18 $^\mathrm{a)}$ &
\text{--} &
\text{--} &
49 \\

NH$_3$ &
102 &
0.41 $^\mathrm{a)}$ &
\text{--} &
\text{--} &
50 \\

CO$_2$ &
86 &
7.5 $^\mathrm{a)}$ &
13 &
87 &
27 \\

C$_3$H$_8$ &
83 &
0.018 $^\mathrm{b)}$ &
6 &
94 &
60 \\

H$_2$S &
80 &
1.8 $^\mathrm{a)}$ &
47 &
53 &
48 \\

C$_2$H$_6$ &
68 &
0.85 $^\mathrm{a)}$ &
5 &
95 &
18 \\

CH$_4$ &
36 &
0.43 $^\mathrm{a)}$ &
54 &
46 &
70 \\

O$_2$ &
30 &
2.3 $^\mathrm{a)}$ &
99.7 &
0.3 &
100 \\

CO &
28 &
3.1 $^\mathrm{a)}$ &
70 &
30 &
72 \\

N$_2$ &
26 &
0.089 $^\mathrm{b)}$ &
63 &
37 &
79 \\

\end{tabular}
\label{tab:results}
\end{center}
\end{table*}

\begin{center}
\begin{figure*}
\includegraphics[width=1.0\textwidth]{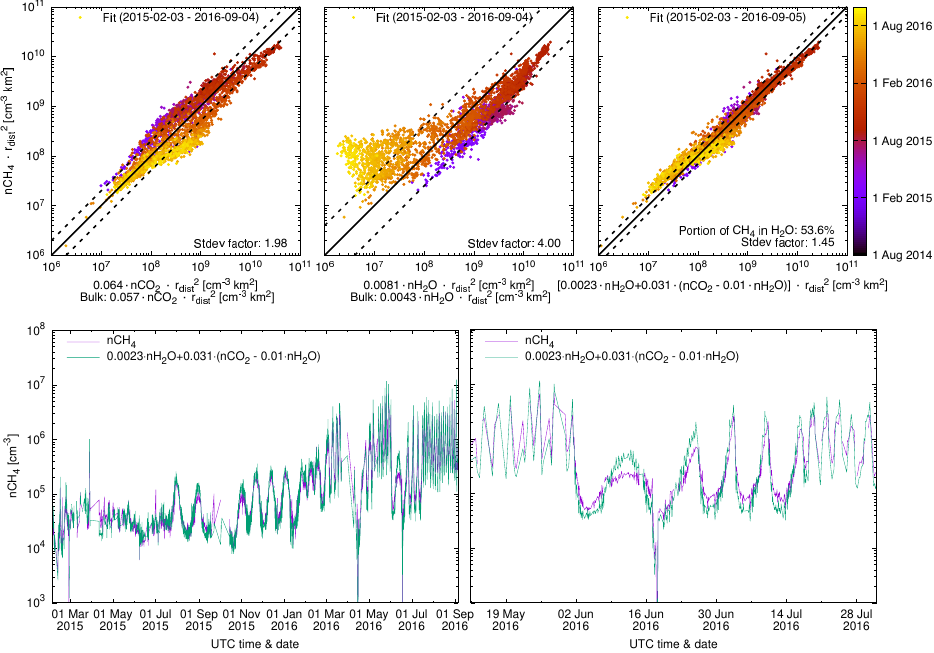}
\caption{Top row: modified densities of CH$_4$ versus CO$_2$ (left), H$_2$O (middle), and a linear combination of both H$_2$O and CO$_2$ (right). Modified density implies multiplication by the cometocentric distance r$_\textrm{dist}$ squared to remove, to first order, the distance-dependence of Rosetta from the comet. The 1\% portion of the CO$_2$ associated with H$_2$O has been taken into account in the right panel. The corresponding coefficients were obtained by fitting the log of the data to assign equal weights to the modified densities which cover several orders of magnitude. The results (e.g., n$_\mathrm{CH_4}$=0.0081$\cdot$n$_\mathrm{H_2O}$) can be compared to the reported bulk abundances (e.g., n$_\mathrm{CH_4}$=0.0043$\cdot$n$_\mathrm{H_2O}$, see text). The 1-standard deviation bounds are provided by the dashed lines (obtained by multiplying and dividing the diagonal solid black line by the indicated factor, respectively). Based on the fitted coefficients from the top right panel it is possible to estimate the portion of the CH$_4$ that is co-desorbed with H$_2$O (54\%, listed in the panel) versus CO$_2$ (see also main text).\\
Bottom row, left panel: CH$_4$ density measured versus the linear combination from the top right panel, i.e. n$_\mathrm{CH_4}$=0.0023$\cdot$n$_\mathrm{H_2O}$+0.031$\cdot$(n$_\mathrm{CO_2}$-0.01$\cdot$n$_\mathrm{H_2O}$) for the early February 2015 though perihelion to September 2016 time period. Right panel: zoom-in to the close ellipses in summer 2016 for comparison to the right panel in Fig.~\ref{fig:H2OCO2}.}
\label{fig:CH4}
\end{figure*}
\end{center}

Fig.~\ref{fig:CH4} shows CH$_4$ in comparison to H$_2$O and CO$_2$. The top left panel presents the abundance of CH$_4$ with respect to carbon dioxide after multiplication with the cometocentric distance squared to account for the variation on cometocentric distance. The second panel shows the same, but relative to water. These modified densities span over several orders of magnitude, therefore the correlation coefficients, listed in the x-axis labels, have been fitted after taking the logarithm of the corresponding values to assign equal weights across the decades. The corresponding results were nCH$_4$=0.064$\cdot$nCO$_2$ and nCH$_4$=0.0081$\cdot$nH$_2$O. A perfect correlation would align the points on the diagonal solid black line. The 1-standard deviation bounds (dashed lines) are offset from the diagonal by the indicated factor. 

The coefficients can also be compared to the corresponding bulk abundances from Table~\ref{tab:results}: nCH$_4$=0.0043$\cdot$nH$_2$O, or when taking CO$_2$/H$_2$O=0.075 into account, follows nCH$_4$=0.057$\cdot$nCO$_2$. The bulk abundance values (from Table~\ref{tab:results} and listed below the x-axis labels in the top row of Fig.~\ref{fig:CH4}), however, are dominated by the measurements during the most active phase of the mission, independent whether the integrated outgassing over the whole mission was considered \citep{Lauter2020} or a suitable period pre-perihelion was investigated \citep{Rubin2019a}. As a result, the red points reflecting the bulk abundances (August 2015, near-perihelion period in the upper right corners) are located below the diagonal line in the case of CO$_2$ (left panel) and H$_2$O (middle panel). The third panel in the top row shows methane versus a linear combination of water and carbon dioxide, taking into account that 1\% CO$_2$ is associated with H$_2$O. Firstly, it can be seen that a linear combination of the H$_2$O and CO$_2$ densities rather well reproduces the measured CH$_4$ density: the data points are well aligned diagonally and the corresponding standard deviation is smallest for the right panel, i.e., the linear combination reproduces the majority of the measurements within 45\% accuracy (within a factor of 1.45 from the diagonal line).

The result of this fitting process is also observed in the bottom two panels: on the left, the reconstructed [n$_\mathrm{CH_4}$=0.0023$\cdot$n$_\mathrm{H_2O}$+0.031$\cdot$(n$_\mathrm{CO_2}$-0.01$\cdot$n$_\mathrm{H_2O}$)] versus the measured  CH$_4$ density is plotted for the whole period investigated here. No modification by the cometocentric distance squared has been applied. The panel on the bottom right shows a zoom into the May to July 2016 period for comparison to the right panel of Fig.~\ref{fig:H2OCO2}. While the match is not perfect, the behavior is well reproduced despite the strong differences between the two host species CO$_2$ and H$_2$O. Also, we have to keep in mind that the whole 19~months period has been taken into account in the fitting process and not just the 3~months shown in the right panel. In our approach, however, we wanted to investigate the different volatiles' behavior throughout the comet's orbit and we hence refrain from fitting individual time periods.

The obtained correlation coefficients are now used to assign the partitions of CH$_4$ related to H$_2$O versus CO$_2$, respectively, when again taking the bulk value of CO$_2$/H$_2$O=0.075 into account: from 0.0023/(0.0023+0.031$\cdot$0.065)=0.54 follows that methane can be associated in about equal portions to water and carbon dioxide. For comparison, if we assume that no CO$_2$ is trapped in or underneath water, follows 0.0023/(0.0023+0.031$\cdot$0.075)=0.50, which yields a very similar conclusion as indicated already in section~\ref{subsec:H2OCO2}. The amount of CO$_2$ that is trapped in or underneath H$_2$O is hence of secondary importance and does not substantially change our results for the minor species.

\begin{center}
\begin{figure*}
\includegraphics[width=1.0\textwidth]{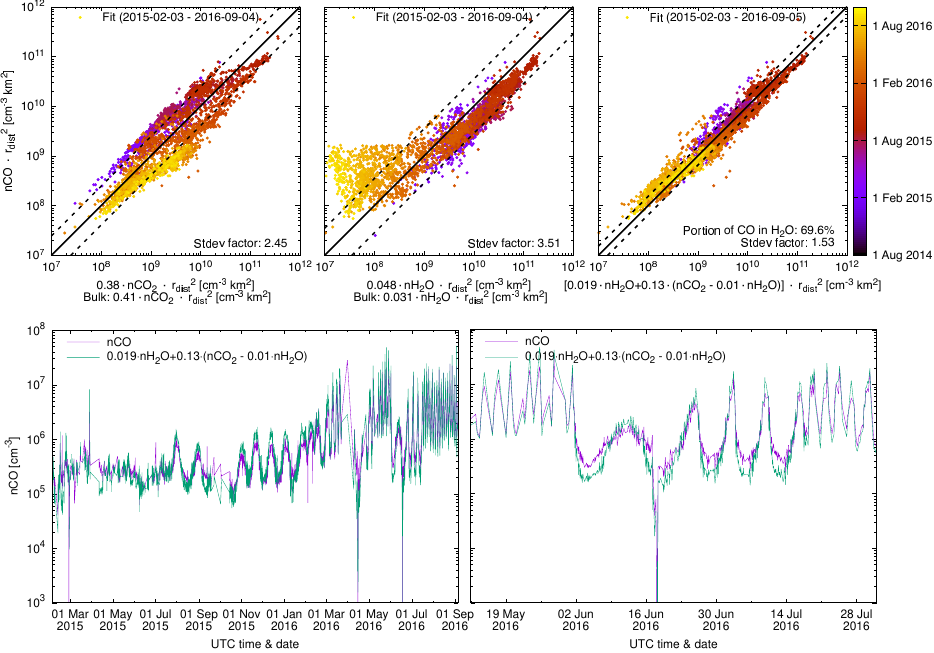}
\caption{Same as Fig.~\ref{fig:CH4} but for CO instead of CH$_4$.}
\label{fig:CO}
\end{figure*}
\end{center}

A similar picture is obtained for CO in Fig.~\ref{fig:CO}. Also here, the modified CO density can be expressed rather well by a linear combination of the modified H$_2$O and CO$_2$ densities. The results reveal a somewhat larger portion of the carbon monoxide, of about 70\%, being associated with water compared to 30\% with carbon dioxide, respectively.

Another molecule we investigated is molecular oxygen, which is of particular interest as it represents one of the most surprising findings in comet 67P \citep{Bieler2015a,Fulle2016}. O$_2$ shows a strong correlation with H$_2$O, which is also reproduced in Fig.~\ref{fig:O2}. Only very little difference is seen between the middle and right panels. Nevertheless, \cite{Luspay2022} showed that a small fraction of the O$_2$ is not correlated to H$_2$O. This is also reproduced in our analysis, i.e., the small portion associated with carbon dioxide can only be observed late in the Rosetta mission, when CO$_2$ was the dominant species in the coma (cf. Fig.~\ref{fig:H2OCO2}). 

Our results for N$_2$, another highly volatile molecule, are provided in Fig.~\ref{fig:N2} and Table~\ref{tab:results}. It should be noted that mass-spectrometric interferences, i.e., the close mass of CO ($\SI{27.9944}{\atomicmassunit\per\elementarycharge}$) to the much lower abundant N$_2$ ($\SI{28.0056}{\atomicmassunit\per\elementarycharge}$), complicates the analysis of the data and increases the uncertainty due to  an additionally required peak fitting procedure (see section~\ref{app:ROSINA} in the appendix). Hence our results for N$_2$ include more scatter and fitting errors than the other species investigated in this work. This is, in parts, also reflected in the larger standard deviation we derived. Still, N$_2$ bears some resemblance to CO, with the larger part associated with H$_2$O and a smaller fraction with CO$_2$.

Ethane, presented in Fig.~\ref{fig:C2H6} \citep[see also][]{Luspay2019}, and propane, presented in Fig.~\ref{fig:C3H8}, are both well correlated to CO$_2$, even though some degree of dichotomy is observed for the time period analyzed, i.e, the early mission values (purple) trend below the diagonal line and later values (yellow) above. This may be a temperature-related fractionation effect, given that the sublimation temperatures of ethane and propane approach the temperature of CO$_2$, cf. Table~\ref{tab:results}. Finally, the same analysis has also been carried out for hydrogen sulfide, H$_2$S in Fig.~\ref{fig:H2S}, another species with sublimation temperature close to that of CO$_2$.

As suggested for the case of ethane, propane, and hydrogen sulfide, the picture for the species of lower volatility changes, i.e., when the species' sublimation temperature rises above the one for carbon dioxide. This shall be discussed in the following section.

\subsection{Species with volatility between CO$_2$ and H$_2$O}\label{sec:lowvolatility}
As shown in section~\ref{sec:TPD}, species of low volatility seldomly co-desorb with species of higher volatility, for instance H$_2$O in CO (cf. Tabs.~\ref{tab:TPD_Kouchi} and \ref{tab:TPD_Gudipati}). Consequently, species like CH$_3$OH are unlikely to co-desorb in abundant amounts in CO$_2$ given the substantially higher pure ice sublimation temperature, i.e., $\SI{142}{\kelvin}$ versus $\SI{86}{\kelvin}$, respectively \citep{Fray2009}. On the other hand, CH$_3$OH does co-desorb with H$_2$O \citep{Martin2014}.

\begin{center}
\begin{figure*}
\includegraphics[width=1.0\textwidth]{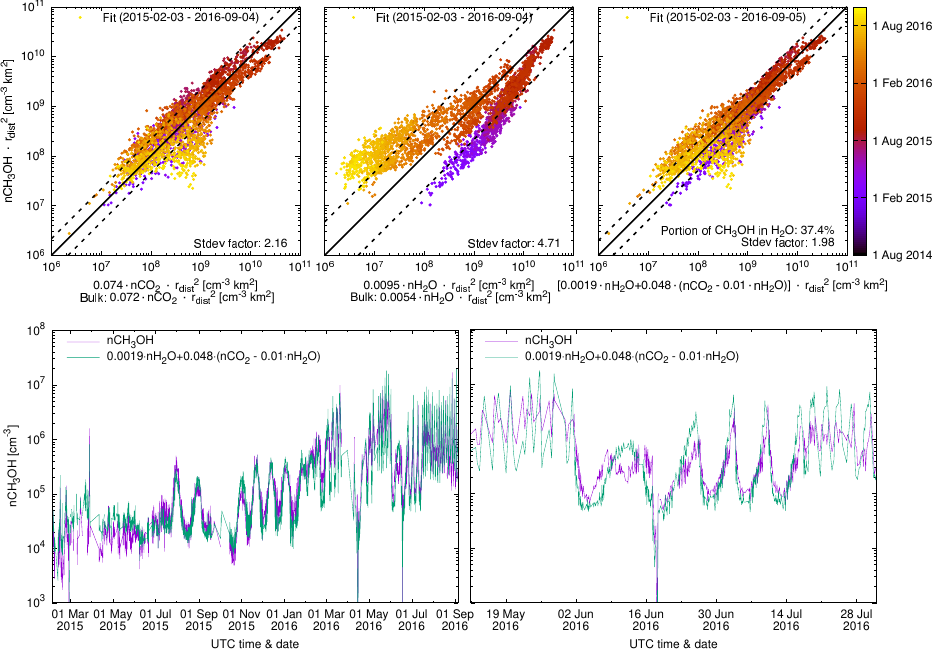}
\caption{Same as Fig.~\ref{fig:CH4} but for CH$_3$OH instead of CH$_4$.}
\label{fig:CH3OH}
\end{figure*}
\end{center}

Fig.~\ref{fig:CH3OH} shows the case of methanol. The top left panel suggests a better correlation of CH$_3$OH with CO$_2$ than H$_2$O (middle panel). The linear combination reveals that about 40\% of the methanol would be associated with water. The standard deviation is larger compared to the highly volatile molecules (cf. section~\ref{subsec:highvolatility}). Furthermore, the bottom right panel shows that methanol follows much closer the water (cf. Fig.~\ref{fig:H2OCO2}) during the low phase angle excursion in the first half of June 2016. This is in contradiction to the reconstructed density from the coefficients obtained from the linear combination of water and carbon dioxide from the top right panel. Additionally, the bottom right panel also reveals systematically overestimated reconstructed densities compared to the measured methanol densities during the April to June 2016 time frame. In addition, the top right panel also reveals some degree of pre-/post-perihelion dichotomy, although less pronounced than for ethane (Fig.~\ref{fig:C2H6}). In accordance with the discussion above, methanol rather may be present in its separate ice phase, or correlated with dust grains that contain volatiles, on top of the portion that is trapped in the lesser volatile water. 

\begin{center}
\begin{figure*}
\includegraphics[width=1.0\textwidth]{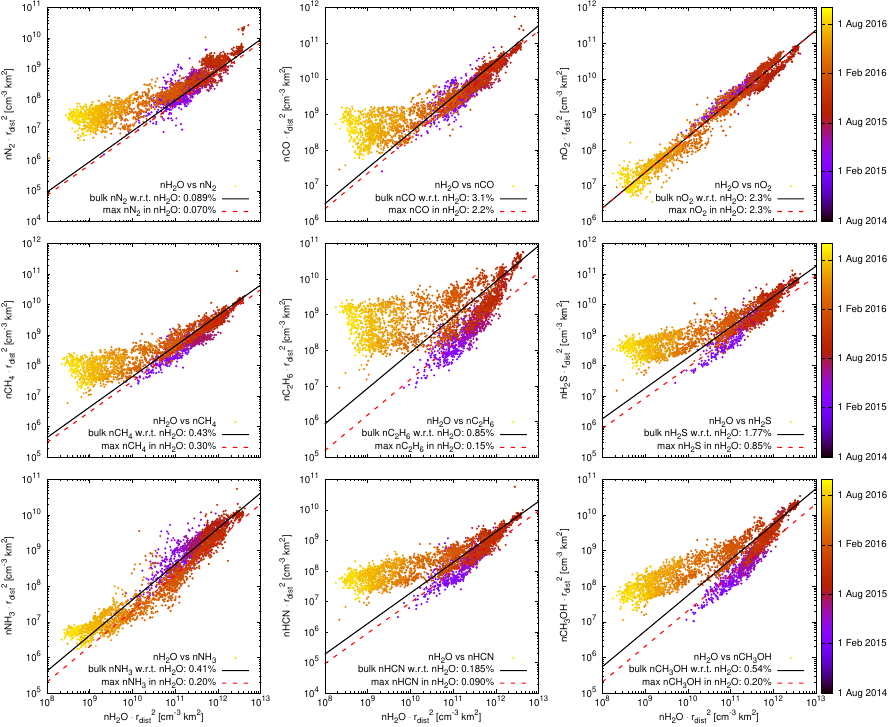}
\caption{A set of cometary molecule densities plotted versus water between February 2015 and September 2016 after multiplication by the cometocentric distance squared (similar to the left panel of Fig~\ref{fig:H2OCO2}). The black line refers to the bulk abundance with respect to water which tends to be close to the peak outgassing period around perihelion (red points). The red dashed line corresponds to an estimated maximum fraction of a given species that is co-released with H$_2$O (see text).
}
\label{fig:MaxTrapH2O}
\end{figure*}
\end{center}

Based on data presented here, it is, however, possible to estimate the maximum amount of methanol that is co-released with water. This is based on similar considerations as applied for the CO$_2$ trapped in or underneath H$_2$O and shown in Fig.~\ref{fig:H2OCO2}. A collection of relative abundances with respect to water for a set of species covering a wide range of sublimation temperatures is presented in Fig~\ref{fig:MaxTrapH2O}. We have also added exploratory lines marking the maximum amounts co-released with water (red dashed lines). For instance, only a small fraction of the C$_2$H$_6$ can be trapped in or underneath H$_2$O, otherwise the measured ethane abundance would have to be larger in the February 2015 time frame. On the other hand, pretty much all O$_2$ can be co-released with H$_2$O as expected from Fig.~\ref{fig:O2}. In the case of methanol, less than 40\% of the density measured at Rosetta may be associated with H$_2$O. \cite{Collings2004} showed that CH$_3$OH has water-like desorption characteristics  (see also \cite{Burke2015} on the influence of CH$_3$OH in water ice. Other relevant TPD experiments by \cite{Martin2014} reveal a complicated picture with shifts in the sublimation temperature as a function of experimental heating rate and CH$_3$OH concentrations in H$_2$O. Depending on the latter, their results also show a variation in the portion of the CH$_3$OH outgassing associated with its own (multilayer) ice phase versus co-release with water, ranging from about 1:1 to almost entirely being associated with water. The situation is hence very complex.

Table~\ref{tab:results} lists these upper limits for a set of volatiles spanning a considerable range in sublimation temperature. As stated above, these limits are derived by visual inspection only and are associated with relative errors estimated to be on the order of $30\%$. However, given the complexity of the cometary activity, these limits can still inform us about the incorporated ices and their outgassing behavior. Consistent with the derived portion associated with the release of water, the upper limits in Table~\ref{tab:results} are higher. Just to provide an example, the portion of C$_2$H$_6$ stored in H$_2$O is small, i.e., 5\% of the total (hence 95\% is associated with CO$_2$), but still smaller than the upper limit of 18\% derived from Fig.~\ref{fig:MaxTrapH2O}.

\subsection{Early mission}\label{subsec:EarlyMission}
\begin{center}
\begin{figure*}
\includegraphics[width=1.0\textwidth]{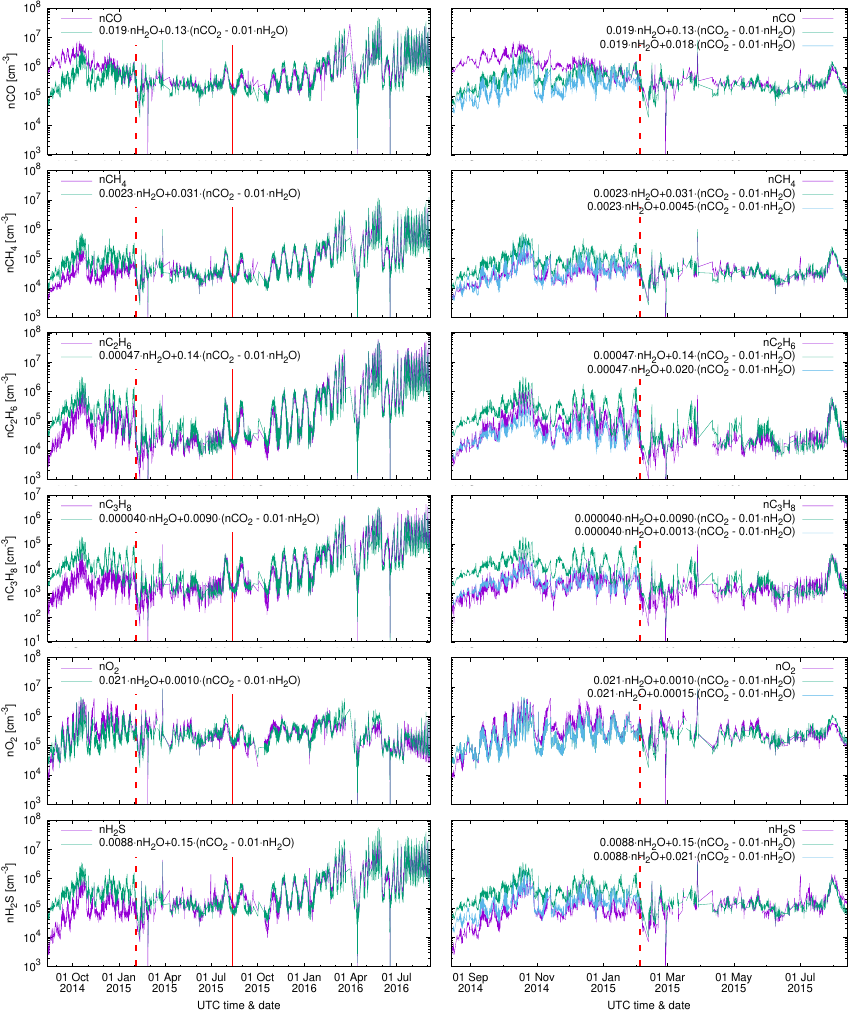}
\caption{Gas densities of highly volatile molecules measured at the Rosetta spacecraft. The left column shows the measured (purple) and reconstructed densities (green) for the entire analyzed period. Perihelion is marked with a red solid line. The right column shows the same up to perihelion and adds a curve representing reduced trapping in CO$_2$ (blue, see also legend and main text) for the time period when the northern hemisphere was more active (up to early February 2015, cf. \protect\cite{Lauter2020}; red dashed line). Top to bottom: CO (cf. Fig.~\ref{fig:CO}), CH$_4$ (cf. Fig.~\ref{fig:CH4}), C$_2$H$_6$ (cf. Fig.~\ref{fig:C2H6}), C$_3$H$_8$ (cf. Fig.~\ref{fig:C3H8}), O$_2$ (cf. Fig.~\ref{fig:O2}), and H$_2$S (cf. Fig.~\ref{fig:H2S}).}
\label{fig:early}
\end{figure*}
\end{center}

In the previous sections, we limited the data to the February 2015 to September 2016 time period when outgassing from the more pristine southern hemisphere dominated. In this section we start mid August 2014, the time of arrival of the Rosetta spacecraft at comet 67P. Fig.~\ref{fig:early} shows the measured (purple) and the reconstructed densities (green) of the highly volatile molecules CO, CH$_4$, C$_2$H$_6$, C$_3$H$_8$, O$_2$, H$_2$S.

The reconstructed densities use the fitted parameters obtained in the previous section~\ref{subsec:highvolatility} (cf. Table~\ref{tab:results}) and were applied to the full time interval. The first key observation, with regard to the early mission phase from mid August 2014 to early February 2015, is that the amount of CO is underestimated. O$_2$, which shows a strong correlation with H$_2$O, remains well reproduced. CH$_4$, C$_2$H$_6$, C$_3$H$_8$, and H$_2$S, which exhibit higher degrees of correlation to CO$_2$, are all overestimated in their reconstructed density. Hence for the early mission phase, we have divided the coefficient reflecting the co-release of the highly volatile molecules in CO$_2$ by a factor 7 and added the corresponding early-mission reconstruction (blue). This correction factor is just an approximation and not an individual fit for each of these highly volatile molecules. Still, the key point here is that the resulting modified reconstruction shows that the early inbound phase is very distinct from the rest of mission, i.e., the CO$_2$ measured early was most likely depleted in trapped highly volatile species. The coefficient representing the co-release with H$_2$O, on the other hand, was left untouched. The implications of these observations are discussed in the next section.

\section{Discussion}\label{sec:Discussion}
This section combines the findings of TPD reference measurements from the literature and our gas coma observations with the goal to study the outgassing behavior of a comet like 67P and to improve our understanding of the ices inside its nucleus. Furthermore, we are interested in the potential differences between families of comets resulting from their distinct dynamical history.

\subsection{Measurement limitations}
Before the discussion of the data presented in the previous section and the comparison to relevant laboratory measurements, there are some limitations to be mentioned. For instance, the relative abundances of the different volatiles used for the TPD measurements do not exactly match the situation of 67P. As mentioned earlier, we investigated the local gas densities at Rosetta and do not directly probe the ices inside the nucleus. We did not include any corrections to the measured gas densities due to gas velocities varying with heliocentric distance, among species, and sources on the nucleus of 67P. Close to perihelion, when the near-nucleus gas densities are highest, the different gas species may be collisionally coupled while this effect plays only a minor role when the comet is far from the sun \citep{Tenishev2008}. Also, our modified gas density assumes a simple r$_\mathrm{dist}^{-2}$ density dependence on the cometocentric distance which can only serve as an approximation.

\subsection{Trapping of volatiles in H$_2$O and CO$_2$}\label{sec:TrappingH2OCO2}
Firstly, we are interested in the behaviour of species of high volatility and their relation to the two main species, CO$_2$ and H$_2$O. As reported by \cite{Gasc2017} and \cite{Luspay2019} and shown in Fig.~\ref{fig:MaxTrapH2O}, no clear distinction between polar and apolar molecules can be made for comet 67P. For instance, H$_2$O (polar) correlates well with O$_2$ (apolar) and NH$_3$ (polar) but not with C$_2$H$_6$ (apolar) and CH$_3$OH (polar). Furthermore, no clear separation according to the pure ice sublimation temperature can be observed, for example, CH$_4$ can be attributed in similar portions to CO$_2$ and H$_2$O, whereas C$_2$H$_6$ almost entirely co-desorbs with CO$_2$. Of course, the situation is more complicated as some of these species may also be linked because of chemical formation while others are not \citep{Herbst2009}.

On the other hand, the local abundance of highly volatile molecules in the coma of 67P, i.e., with a pure ice sublimation temperature below CO$_2$, can be described as a linear combination of the local H$_2$O and CO$_2$ densities. Ice phases dominated by highly volatile species cannot be excluded, however, they are not required to explain the measured abundances in the coma. The majority of highly volatile molecules are associated with H$_2$O and CO$_2$ in varying proportions. The observed slopes of the different highly volatile molecules reported by \cite{Gasc2017}, i.e, the heliocentric distance dependence of their outgassing (cf. section~\ref{subsec:67P}), can be confirmed to be the result of the slopes of the two main species, H$_2$O and CO$_2$, and the corresponding associated fractions.

The in situ coma measurements obtained with ROSINA cannot resolve the release of gases in the temperature range of the molecular volcano compared to the main H$_2$O outgassing peak, which occurs at slightly higher temperatures. As shown in both Figs.~\ref{fig:TPD_Kouchi} and \ref{fig:TPD_Gudipati}, when just focusing on the temperature range of the molecular volcano, all trapped volatiles are released together during the amorphous to crystalline ice phase transition. For instance, O$_2$ would rather be correlated with CO$_2$ than H$_2$O. However, we observed only very little of the O$_2$ to be associated with CO$_2$ (Table~\ref{tab:results}). This indicates that the co-release of O$_2$ and CO$_2$ during the molecular volcano has only a minor influence on the observed correlations, i.e, was accounted for by attributing 13\% of the total CO$_2$ to the co-release with water and subtracting this in our fitting procedure (section~\ref{subsec:H2OCO2}).

Species of lower volatility than CO$_2$, on the other hand, may be trapped in or underneath H$_2$O and be present in their own (multilayer) ice phase. Our analysis yields upper limit estimates for the fraction co-released with water and the results are consistent with the coefficients associated with co-release with water from the linear fits (Table~\ref{tab:results}). There may also be a transition regime for species with sublimation temperature close to CO$_2$ as they may co-release with both water and carbon dioxide as well as forming their own phase. Possible candidates are C$_2$H$_6$ (cf. Fig.~\ref{fig:C2H6}), C$_3$H$_8$ (cf. Fig.~\ref{fig:C3H8}), and H$_2$S (cf. Fig.~\ref{fig:H2S}) which show some degree of pre- to post-perihelion difference (cf. section~\ref{subsec:highvolatility}).

\subsection{Loss of highly volatile species through thermal processing}
In the previous section, we showed that the measured abundances of highly volatile molecules can be explained by trapping in or underneath both water and carbon dioxide. The presented fits do not include an additional component representing the ice phase dominated by the highly volatile molecules CH$_4$, CO, etc. The first row in both Tabs.~\ref{tab:TPD_Kouchi} and \ref{tab:TPD_Gudipati} lists the fraction of the volatiles released in the temperature range $\SI{<70}{\kelvin}$. According to these results, O$_2$, CO, and CH$_4$ are all heavily affected and to a lesser extent also N$_2$. Substantial amounts of these species would hence have been lost from the cometary nucleus of a JFC like 67P prior to arrival in the inner solar system. For the presented TPD measurements, with a trapping temperature of $\SI{10}{\kelvin}$, the results are summarized in Table~\ref{tab:lost}. With subsequent heating to $\SI{\sim70}{\kelvin}$ after trapping, a comet like 67P would have lost up to 70\% of its initial CO content.

Nevertheless, our knowledge of the initial trapping temperature or range thereof is very limited. The lost fraction of highly volatile species very much depends on the conditions during trapping. Additional constraints on the temperature during the formation of the material incorporated into 67P may thus be required. These can be obtained from independent considerations: for instance, a chemical reaction model for dark clouds by \cite{Taquet2016} suggests that temperatures of approximately $\SI{20}{\kelvin}$, which is somewhat elevated compared to typical dark cloud temperatures (5 -- $\SI{10}{\kelvin}$), may be required to enhance the O$_2$ abundance to be comparable to the one found in 67P \citep{Bieler2015a}. In summary, it remains unclear how much of the highly volatile species were initially trapped and subsequently lost during the dynamical history of the comet.

\begin{table}
\begin{center}
\caption{Lost fraction [\%] of highly volatile molecules, if originally trapped at $\SI{10}{\kelvin}$ and subsequently heated to $\SI{70}{\kelvin}$, e.g., for 67P during the Centaur phase \protect\cite[$\SI{\sim7}{\mega\year}$ at $\SI{7}{\astronomicalunit}$;][]{Guilbert2016}. CH$_4$ was taken from Table~\ref{tab:TPD_Kouchi} and O$_2$, N$_2$, and CO from Table~\ref{tab:TPD_Gudipati}.}
\begin{tabular}{ c  c }
Species & Lost fraction \\ \hline
CH$_4$ & 61\% \\
O$_2$ & 82\% \\
CO & 70\% \\
N$_2$ & 39\%

\end{tabular}
\label{tab:lost}
\end{center}
\end{table}

Thermal modelling, e.g., by \cite{DeSanctis2001} and \cite{Parhi2023}, shows that substantial amounts of CO-dominated ices will be lost from cometary nuclei in the Kuiper belt at distances beyond $\SI{40}{\astronomicalunit}$. Loss of the ice phase of highly volatile species may already occur earlier, for instance due to collisional heating during the formation of the Kuiper belt \citep{Jutzi2020}. Thermal processing of icy primordial disc objects also took place before scattering to the different cometary reservoirs occurred as a result of giant planet migration \citep{Gomes2005}. For instance, \cite{Davidsson2021b} found that objects in the size range of 4-$\SI{200}{\kilo\meter}$ would lose their CO-dominated ice content within 0.1-$\SI{10}{\mega\year}$ due to heating by the protosun and long-lived radionuclides. For small objects, the latter may be less efficient, as shown by \cite{Mousis2017}. Furthermore, an early loss of the N$_2$- and CO-dominated ices is in line with the finding that the N$_2$/CO ratio is independent of the number of inner solar system crossings \citep{Anderson2023} given that the outgassing of both highly volatile molecules is governed by the lesser volatile CO$_2$ and H$_2$O. For this, we assume that the CO$_2$/H$_2$O ratio does not change significantly from one perihelion passage to the next. For the same reason, however, the N$_2$/CO ratio may change over a single orbit due to the change of the CO$_2$/H$_2$O ratio as a function of heliocentric distance. This may then also explain the variation in the N$_2$/CO ratio observed in 67P \citep{Rubin2015b,Rubin2019a}. 

The thermal history of a comet similar to 67P may lead to additional depletion of highly volatile species. Here, it is important to consider the dynamic history of 67P \citep{Maquet2015}. Before becoming a JFC, 67P may have spent several million years as a Centaur at an intermediate heliocentric distance around $\SI{7}{\astronomicalunit}$ and undergone internal heating to temperatures of up to $\SI{70}{\kelvin}$ \citep{Guilbert2016}. This can be compared to comet 29P/Schwassmann-Wachmann~1, a Centaur on its way to become a JFC \citep{Sarid2019}, which shows substantial and routinely detected CO-outgassing \citep{Crovisier1995}. As a consequence, CO may have been another major species in cometary ices, along with H$_2$O and CO$_2$, before substantial amounts were lost due to the heating processes. Accordingly, some of the comets originating from the Oort cloud, e.g., 1P/Halley \citep{Eberhardt1999}, exhibit higher relative CO abundances compared to CO$_2$. Also long-period comet (LPC) C/2016 R2 (PanSTARRS), a returning object originating from the Oort cloud, showed both strong N$_2^+$ and CO$^+$ emission bands \citep{Opitom2019b}. In the ISM, for comparison, CO has been shown to be high in abundance, both in the gas and the ice phases \citep{Boogert2015,Zamirri2018}. As discussed earlier in section~\ref{sec:TPD} and shown by \cite{Parhi2023}, if ample time is available, the same highly volatile species may be also lost at much lower temperatures. We therefore state that the $\SI{70}{\kelvin}$ internal temperature mentioned above is just an upper limit.

These conclusions are compared to those of \cite{AHearn2012}, who investigated the CO/CO$_2$/H$_2$O composition of comets. Their survey includes, aside from JFCs, also LPCs and Halley-type comets (HTCs). The collection of observations shows a large variation in relative abundances of the three species within each of the families of comets, which is hence most likely not the result of their different dynamical origin. Little systematic difference between the families of comets were observed, i.e., between JFCs, which reside as a Centaur for an extended period at intermediate distance \citep{Guilbert2016}, versus LPCs and HTCs. The CO/H$_2$O ratio did, however, trend to somewhat higher ratios for LPCs when compared to JFCs. However, the two orders of magnitude variation in CO/H$_2$O in the sample of comets analyzed by \cite{AHearn2012} complicates the identification of clear trends. 

\begin{center}
\begin{figure}
\includegraphics[width=1.0\columnwidth]{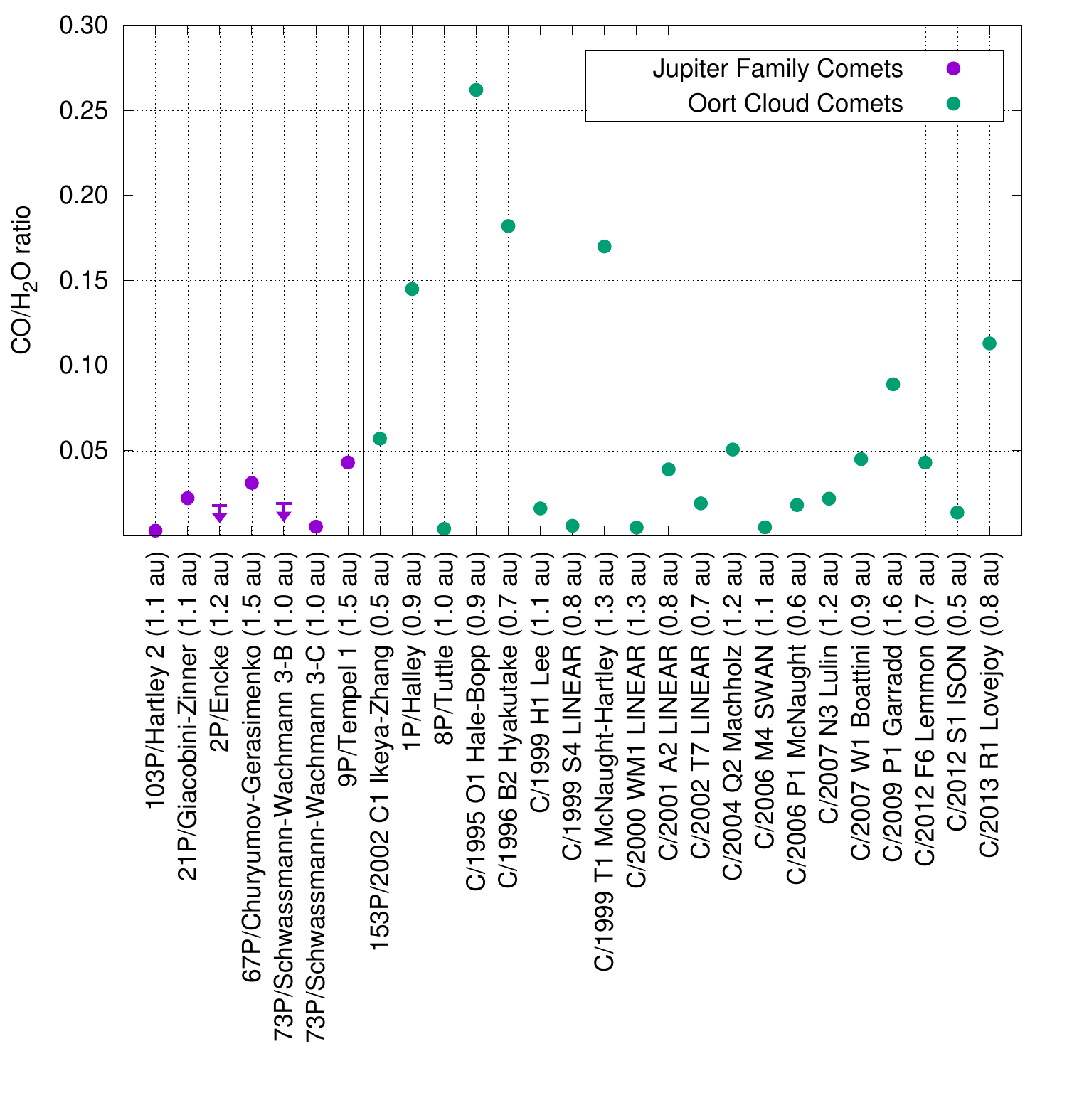}
\caption{CO/H$_2$O ratio measured in range of JFCs and OCCs based on infrared spectroscopic observations \citep{DelloRusso2016a}, amended by the two comets for which mass spectrometric measurements are available , i.e., 1P/Halley and 67P, \protect\cite{Eberhardt1999} and \protect\cite{Rubin2019a}, respectively. The label also contains the heliocentric distance of the corresponding observation.}
\label{fig:COinComets}
\end{figure}
\end{center}

To further delve into this issue, Fig.~\ref{fig:COinComets} shows a collection of measured CO abundances with respect to H$_2$O in both JFCs and OCCs based on infrared spectroscopic \citep{DelloRusso2016a} and in situ mass spectrometric observations \citep{Eberhardt1999,Rubin2019a}. While the ratios are limited to several percent in JFCs, the variation of CO/H$_2$O in OCCs is substantially larger. The latter family includes a subset with much higher CO/H$_2$O ratios. This may be the result of the aforementioned transition through the Centaur stage and other thermal processes related to their dynamical history and possibly the size of the objects, e.g., heating due to the collisional cascade in the primordial disk \citep{Davidsson2023}. Consequently, our results support the suggestion by \cite{Gasc2017} and \cite{Davidsson2021b} that highly volatile species, such as CO, are trapped in (or underneath) less volatile species like H$_2$O and CO$_2$ to remain present in comet 67P's nucleus.

There still remain questions regarding the evolution of comets and associated loss of highly volatile species, for instance, \cite{Harrington-Pinto2022} reported lower CO/CO$_2$ ratios in Dynamically New Comets (DNCs) compared to the more processed OCCs. Furthermore, this ratio may even increase with the dynamical age of the comet \citep{AHearn2012}. A possible explanation is the irradiation of a DNC by GCRs leading to a depletion of the CO in the top layer \citep{Gronoff2020,Maggiolo2020} which is then shed during the first apparition.

\cite{Luspay2022} reported that the original O$_2$, regardless of its origin in the ISM or in the PSN, may have been incorporated into the comet at abundances lower than what is found today. They investigated DFMS data obtained during select time periods above 67P's southern hemisphere at heliocentric distances $\SI{>2.4}{\astronomicalunit}$. In their scenario, evolutionary processing involving secondary trapping in the underlying H$_2$O matrix leads to the elevated ratio observed by Rosetta. This is somewhat different from the scenario discussed here, which is based on data from almost the entire mission including perihelion and the northern latitudes. Depending on the formation temperature, it is likely that highly volatile species were more abundant in comet 67P after its formation. Also, our scenario to explain the O$_2$ observations at 67P does not include the formation of a secondary reservoir but the primordial trapping of O$_2$ in mostly H$_2$O, with a small fraction in CO$_2$, similar to the corresponding TPD experiments by \cite{Gudipati2023}.

Another key element is nitrogen: cometary volatiles seem to lack a substantial fraction of elemental nitrogen with respect to solar relative abundances of other volatile elements like C and O \citep{Geiss1988}. Several processes have been discussed in the literature, for instance the incorporation of N-bearing species, predominantly NH$_3$, into ammonium salts of lower volatility \citep{Altwegg2020a}. As previously discussed in the literature \citep[e.g.,][]{Geiss1988}, the loss of N$_2$ may also be responsible for a part of the missing nitrogen. Our finding that the measured molecular nitrogen abundance in 67P can be reproduced without including an N$_2$-dominated phase further supports this scenario. On the other hand, the amount of N$_2$ lost is limited (Table~\ref{tab:lost}) and not sufficient to explain the nitrogen deficiency alone.

Table~\ref{tab:lost} also suggests that loss of the ice phase dominated by the highly volatile species is expected to change their ratio as the nucleus undergoes collisional and thermal heating processes. For instance, the N$_2$/CO ratio may be evolutionary altered which increases the uncertainty when deriving the formation temperature of the incorporated ices \citep{Rubin2015b}.

An interesting observation in this regard is the apparently distinct behavior of the hydrocarbon species investigated in this work. Methane, ethane and propane have pure ice sublimation temperatures below carbon dioxide (Tab~\ref{tab:results}). Ethane in Fig.~\ref{fig:C2H6} and propane  in Fig.~\ref{fig:C3H8} are well correlated to CO$_2$, as opposed to H$_2$O, which is quite distinct from the 1:1 partitioning found for methane according to Fig.~\ref{fig:CH4}. This, however, may be explained by the loss of the methane-dominated phase, which may amount to approximately 60\% of the total according to Table~\ref{tab:lost}. Therefore, when taking into account the lost portion, the fraction of the total CH$_4$ which is associated with H$_2$O would reduce to about 19\% (cf. Table~\ref{tab:TPD_Kouchi}), which is closer to the partitioning of ethane and propane ($\sim$5-6\% associated with H$_2$O). While we lack reference TPD measurements for ethane and propane in a mixture of CO$_2$ and H$_2$O, their pure ice phase sublimation temperature is much closer to the temperature the comet may have witnessed in its past or even close to CO$_2$ and hence a substantial fraction of the ethane-dominated and propane-dominated ice may still be present inside the nucleus. The preferential loss of methane over ethane is supported by \cite{Parhi2023}, who applied a comet evolution model to compute depletion times for different volatiles in Kuiper belt objects (KBOs). Further relevant are the results by \cite{Schuhmann2019}, showing that methane and ethane have very similar abundances in the coma of 67P, while other, heavier aliphatic hydrocarbons such as propane showed abundances decreasing with the mass of the molecule. A similar picture was also obtained near perihelion by \cite{Hanni2022}. At that time, a substantial amount of cometary dust was present in the coma. 

One caveat is that the sublimation temperature of ethane is also close to carbon dioxide (Table~\ref{tab:results}). This may result in a correlation between the two without causation. In fact, the top right panel in Fig.~\ref{fig:C2H6} does reveal some degree of pre-to-post-perihelion difference, similar to H$_2$S (Fig.~\ref{fig:H2S}) and C$_3$H$_8$ (Fig.~\ref{fig:C3H8}), both species with sublimation temperatures even closer to CO$_2$. 

In summary, the presence of highly volatile molecules in comets is, hence, a less stringent limit on the degree of internal heating that could have occurred. Not suitable to recover the thermal history of the comet may be the ortho-to-para ratio (OPR) in water and other molecules: previously, the OPR has been employed to obtain the formation temperature of the ices incorporated into a comet. However, recent work has questioned this connection by showing that the OPR measured in, e.g., coma H$_2$O and NH$_2$ is independent of the original ice formation process \citep{Cheng2022,Shinnaka2016c,Faggi2018}.

Other constraints are required, for instance, the limited deuterium-hydrogen isotope exchange reactions, which become significant above temperatures of $\SI{70}{\kelvin}$ on time-scales of $\SI{1e4}{\year}$ \citep{Lamberts2015}. For a comet like 67P, with very different D/H ratios in water when derived from D/H=0.5$\cdot$HDO/H$_2$O versus D/H=2$\cdot$D$_2$O/HDO \citep{Altwegg2017a}, this indicates that its internal temperature remained well below $\SI{70}{\kelvin}$ for most of its lifetime. 

Another constraint is obtained from the chemical model by \cite{Taquet2016}, who suggested a formation temperature on the order of $\SI{\sim20}{\kelvin}$ based on the O$_2$/H$_2$O ratio measured in 67P \citep{Bieler2015a}.

\subsection{Comparison with TPD measurements}
After neglecting the outgassing in the $<\SI{70}{\kelvin}$ temperature range (cf. Table~\ref{tab:TPD_ranges}), the obtained correlation factors can be compared to the TPD measurements in Tabs.~\ref{tab:TPD_Kouchi} and \ref{tab:TPD_Gudipati}. We thus estimate the portions of N$_2$, O$_2$, CO, and CH$_4$, respectively, associated with CO$_2$ (70--$\SI{115}{\kelvin}$ temperature range) and H$_2$O ($\SI{>115}{\kelvin}$ temperature range). For instance, based on the TPD measurements by \cite{Kouchi1995} in Table~\ref{tab:TPD_Kouchi}, the portion of the CH$_4$ outgassing associated with CO$_2$ is 0.203/(0.203+0.185)=52\% which leaves 48\% to H$_2$O. The corresponding result from our fitting procedure of the relative abundances in 67P is 46\% of the CH$_4$ associated with CO$_2$ and 54\% to H$_2$O (see Fig.~\ref{fig:CH4}). For CO, and based on the TPD results by \cite{Kouchi1995} in Table~\ref{tab:TPD_Kouchi}, the portion of the outgassing associated with CO$_2$ is 0.172/(0.172+0.071)=71\% and the remaining 29\% are released together with the H$_2$O. When using TPD measurements from \cite{Gudipati2023} in Table~\ref{tab:TPD_Gudipati}, however, 24\% are associated with CO$_2$ and 76\% to H$_2$O. The corresponding results for 67P are 30\% of the CO associated with CO$_2$ and 70\% to H$_2$O (see Fig.~\ref{fig:CO}).

 A collection of these results is presented in Table~\ref{tab:TPD_vs_67P}. The top two rows summarize our results for 67P, using two different CO$_2$/H$_2$O ratios, once from the integrated total relative outgassing during the whole Rosetta mission \citep{Lauter2020} and once measured pre-perihelion just before the peak outgassing period \citep{Rubin2019a}. The bottom two rows show the corresponding TPD measurements \citep{Kouchi1995,Gudipati2023}. It becomes obvious that there is no single TPD measurement that matches the ratios measured for 67P. This, however, is no surprise given that the volatiles measured by ROSINA at 67P originate from various locations which hence cover, among others, a wide variation of local temperatures, ice composition, and depths from which the sublimation occurs \citep{Combi2020}.

\subsection{Impact of hemispherical transport of icy grains}
Once the southern hemisphere became active, a fraction of the grains which contain volatiles were entrained in the gas flow, lifted and transported to the northern hemisphere, and then deposited due to gravity and lower gas drag. The northern hemisphere is thus subject to redeposition of material and is hence less pristine. Such hemispherical transport of icy grains has been observed and is tied to the prominent deposit regions in the northern regions of the comet \citep{Thomas2015,El-Maarry2015,Keller2017}. The southern hemisphere, on the other hand, is dominated by erosion processes which continuously expose fresh material from the comet's interior \citep{Keller2015a}. 

\cite{Davidsson2021a} showed that H$_2$O ices are likely to survive this journey, as opposed to the CO$_2$-dominated phase, which may be lost even from large decimeter-sized chunks of ice. ROSINA measurements show that there are still substantial amounts of CO$_2$ present in the coma above the north. Our results thus support that at least parts of the CO$_2$ is trapped in H$_2$O and hence survives the journey to the northern hemisphere deposits where it is then released together with the water, consistent with the model by \cite{Davidsson2021a}.

Another effect that may be contributing to non-zero CO$_2$ abundances everywhere in the coma are inter-particle collisions and surface scattering processes which redistribute molecules around the nucleus, potentially reaching Rosetta even if there is no direct line-of-sight to the corresponding source region. This effect is expected to be more prominent near perihelion, when coma densities were high and collisions more frequent \citep[cf.][]{Combi2012}. However, the lowest CO$_2$/H$_2$O ratios were measured well before perihelion at distances $>\SI{2}{\astronomicalunit}$ (cf. Fig.~\ref{fig:H2OCO2}), showing that collisions were not the main cause for the minimum abundance of CO$_2$ which was present throughout the coma.

\subsection{Seasonal CO$_2$ frost}\label{sec:frostresult}
Our fits from section~\ref{sec:TrappingH2OCO2} can be applied throughout most of the Rosetta mission, covering a wide range in heliocentric distances, except the early period from August 2014 to February 2015. As a result of the comet's substantial obliquity of the spin axis \citep{Sierks2015}, 67P exhibits strong seasonal outgassing \citep{Hassig2015}. During the inbound part of the orbit, the sub-solar latitude gradually moved from the north towards the south (Fig.~\ref{fig:H2OCO2}). \cite{Lauter2020} showed that early in the mission CO$_2$ sublimated mostly from the south, whereas H$_2$O followed the sub-solar latitude and thus originated during that time from the northern summer hemisphere. The CO$_2$, however, appeared to be depleted of highly volatile molecules, possibly due to the sublimation and re-freezing of the CO$_2$ during the previous outbound arc, when the interior of the nucleus was still warmer compared to the southern hemisphere surface which gradually went into shadow. However, the embedded highly volatile species released at the same time would not be re-trapped or frozen-out efficiently and hence be lost. In fact, the VIRTIS instrument directly observed such patches of CO$_2$ frost on 67P's surface as they emerged from local winter \citep{Filacchione2016}. The CO$_2$ frost observed by ROSINA, on the other hand, may not be limited to the surface but also originate from well below. Aside from CO$_2$, VIRTIS observations by \cite{DeSanctis2015} also revealed H$_2$O surface frost acting on a diurnal time scale (cf. section~\ref{subsec:SupportingObs}).

For the rest of the Rosetta mission, CO$_2$ remained outgassing predominantly from the south, in particular also after the outbound equinox in March 2015, despite the sub-solar latitude moving back to the northern latitudes \citep{Lauter2019}.  

When investigating the absolute production rates, CO$_2$ showed an almost stable, possibly even slightly decreasing gas production inbound from $\SI{>3.5}{\astronomicalunit}$ to $\SI{\sim2.5}{\astronomicalunit}$ \citep{Lauter2020,Combi2020}. Again, this may be the result of the CO$_2$ frost, released from greater depth and re-freezing closer to the colder surface during the previous outbound journey. After aphelion and inbound during the next orbit, the heat wave starts again from the surface inwards and releases the seasonal CO$_2$ frost, which is now located closer to the surface and devoid of trapped highly volatile species. This, in essence, enhances the comet's early CO$_2$ activity. Only inside $\SI{\sim2.5}{\astronomicalunit}$ the CO$_2$ production clearly increased with decreasing heliocentric distance. This then marks again the point when outgassing of fresh and more pristine CO$_2$, with all of the trapped volatiles, starts.

If we assume that all the CO$_2$ emitted from the southern hemisphere before February 2015, the period for which we found low amounts of highly volatile molecules co-released with CO$_2$, is frost, then we can estimate the amount of CO$_2$ frost stored in 67P. For this purpose, we integrate the southern hemisphere CO$_2$ production from \cite{Lauter2019} during this early inbound phase and compare this to the total CO$_2$ production per orbit. The result yields that $0.68^{+0.25}_{-0.16}$\% of the total CO$_2$ production may be inherited from the previous apparition in the form of frost. Given that the sub-solar latitude is moving farther south afterwards (Fig.~\ref{fig:H2OCO2}), additional CO$_2$-frost may also be released after this point in time. However, its contribution is then hidden in the sublimation of fresh CO$_2$, which still contains all the trapped species and dominates the outgassing. But also before February 2015, a fraction of the outgassing CO$_2$ from the southern hemisphere may be pristine. Therefore, the provided uncertainty reflects the outgassing of CO$_2$ frost ending either one month earlier or later (compared to early February 2015 shown in Fig.~\ref{fig:early}), respectively. 

Despite these uncertainties, frost seems to be an important process involving various species and different time as well as length scales, i.e., volatiles, upon sublimation, may leave the comet or re-freeze on other, colder locations inside the  nucleus or on its surface. It is thus also interesting to compare our results to the recent H$_2$O detection at main belt comet 238P/Read: \cite{Kelley2023} reported a CO$_2$/H$_2$O<0.7\%. The comet's CO$_2$-dominated ice phase may thus have been lost over time or its CO$_2$ activity is quenched. At 67P, we found about 1\% CO$_2$ to be present and associated with H$_2$O. In order to reduce the minimum amount of CO$_2$ with respect to H$_2$O in 67P, the comet would have to undergo water frost cycles that release trapped volatiles similar to the situation discussed for CO$_2$. In the case of comet 238P/Read, there are temporal changes in the observed H$_2$O activity: \cite{Kelley2023} reported that devolatilization of the surface takes place on orbital timescales and the comet's activity is gradually reduced. Frequent H$_2$O frost cycles may hence have occurred in the past or may even be ongoing.

\subsection{The odd case of CO}
CO remains poorly understood during the early mission phase. As opposed to the other highly volatile molecules (cf. section~\ref{sec:frostresult}), the CO density, measured locally at Rosetta, was enhanced and not depleted when the density was reconstructed based on the fractions associated with H$_2$O and CO$_2$. Furthermore, when investigating the absolute production rate, CO was first elevated and showed a decreasing activity inbound from $\SI{>3.5}{\astronomicalunit}$ to $\SI{\sim2.5}{\astronomicalunit}$, even more pronounced than CO$_2$ \citep{Lauter2020,Combi2020}.

It is interesting to note that this peculiar behavior has been observed before, e.g., inbound at just about the same heliocentric distance range at comet Hale-Bopp \citep{Biver2002a}, another returning comet \citep{Bailey1996} with strong seasonal outgassing \citep{Kuhrt1999} associated with a highly oblique spin axis \citep{Jorda1997}.

The enhanced CO outgassing occurs during the same early time period of alleged CO$_2$ frost outgassing (cf. section~\ref{sec:frostresult}). This may indicate a common process, i.e., the release of CO frost, which is hence not associated with H$_2$O and CO$_2$ and thus underestimated in the reconstructed density. However, this would raise some additional questions, in particular associated with the substantially lower sublimation temperature of CO compared to CO$_2$ (cf. Table~\ref{tab:results}). Also CH$_4$, which has a comparably low sublimation temperature and is co-released in similar proportions with H$_2$O and CO$_2$, does not show the same peculiar behavior or at least to a lesser degree.

Another possibility is irradiation of CO$_2$ ice which results in the formation of CO, which in turn results in CO trapped in CO$_2$ and thus follows a strong correlation between the two molecules \citep{Martin2015}. But also this explanation has a number of shortcomings. Seasonal carbon dioxide frost has been seen on the surface of the southern hemisphere of 67P \citep{Filacchione2016}, however, there may be a lot more stored underneath the surface, protected from most photons and energetic particles. Additionally, the southern hemisphere was in winter during the aphelion period, hence a large fraction of its surface witnessed constant night which limits photolysis and largely protects it from radiolysis by solar wind particles. Furthermore, the heliocentric distance was beyond $\SI{4}{\astronomicalunit}$ for most parts of that phase which leads to low photon and solar wind particle fluxes.

In this discussion, however, we cannot delve into the details associated with the retention of these highly volatile molecules. Clearly, further work is required to understand frost cycles involving different species and the associated diffusion through the porous media of the comet's nucleus including release, re-trapping, and re-freezing processes.

\subsection{Future mission outlook}
ESA's new F-class mission Comet Interceptor is designed to be the first spacecraft to fly-by a DNC, i.e., a comet that will enter the inner solar system for the first time \citep{Snodgrass2019}. Such a target is possibly more pristine compared to the comets visited by spacecraft thus far. CO, for instance, could still be a major (host) species. 

Heating and hence loss of highly volatile species through irradiation by the protosun \citep{Davidsson2021b} and by galactic cosmic rays \citep{Gronoff2020} is altering the outermost tens of meters of objects residing in either the Oort Cloud or the Kuiper Belt \citep{Maggiolo2020}. Additionally, heating by catastrophic collisions \citep{Jutzi2020,Davidsson2023} and radiogenic elements \citep{DeSanctis2001} may have taken place already. Putting additional constraints on the degree of these alterations is a major goal of Comet Interceptor. One key measurement of the mission will hence be to obtain relative abundances of highly volatile molecules, such as CO, O$_2$, and CH$_4$, for comparison to the ones observed in different families of comets.

Nevertheless, the situation remains complicated, e.g., a recent survey has shown that DNCs tend to have lower CO/CO$_2$ ratios compared to OCCs \citep{Harrington-Pinto2022}.

Furthermore, our findings may also be relevant for a cryogenic sample return from the surface of a comet. Since the most likely target for such a mission is a JFC, it is estimated that the ice phases of highly volatile species are lost from at least the surface and outgassing layer of the nucleus.

\begin{table*}
\begin{center}
\caption{Relative portion of highly volatile molecules associated with either H$_2$O (release in the $\SI{>115}{\kelvin}$ temperature range) or CO$_2$ (70--$\SI{115}{\kelvin}$) from TPD measurements and observed in 67P. For the latter, the results were derived once using the CO$_2$/H$_2$O ratio from the mission integrated outgassing \protect\citep{Lauter2020} and once from the relative outgassing during the period just before perihelion \protect\citep{Rubin2019a}.}
\begin{tabular} {l | c | c c | c c | c c | c c }

\multicolumn{1}{l |}{} &
\multicolumn{1}{c |}{CO$_2$} &
\multicolumn{2}{c |}{CO} &
\multicolumn{2}{c |}{CH$_4$} &
\multicolumn{2}{c |}{N$_2$} &
\multicolumn{2}{c}{O$_2$} \\

associated species (temperature regime) &
H$_2$O &
H$_2$O &
CO$_2$ &
H$_2$O &
CO$_2$ &
H$_2$O &
CO$_2$ &
H$_2$O &
CO$_2$ \\ \hline \hline

67P for CO$_2$/H$_2$O=0.075 \citep{Lauter2020}&
13\% &
70\% &
30\% &
54\% &
46\% &
63\% &
37\% &
99.7\% &
0.3\% \\

67P for CO$_2$/H$_2$O=0.047 \citep{Rubin2019a} &
21\% &
80\% &
20\% &
66\% &
34\% &
75\% &
25\% &
99.8\% &
0.2\% \\ \hline

TPD \citep{Kouchi1995} &
20\% &
29\% &
71\% &
48\% &
52\% &
-- &
-- &
-- &
\\

TPD \citep{Gudipati2023} &
56\% &
76\% &
24\% &
-- &
-- &
69\% &
31\% &
70\%&
30\% \\

\end{tabular}
\label{tab:TPD_vs_67P}
\end{center}
\end{table*}

\section{Summary \& Conclusions}\label{sec:Conclusions}
In this paper, we analyzed data from the ROSINA mass spectrometer DFMS and pressure sensor COPS, obtained throughout most of the Rosetta mission time, i.e., during a period of more than 24~months. The combined measurements provide the local gas density at the location of the Rosetta spacecraft. From this dataset, we investigated correlations among different volatile species with regard to the two main ice phases in comet 67P, H$_2$O and CO$_2$. A key finding is that the local abundance of the highly volatile molecules, such as CO and CH$_4$, can be reproduced by a linear combination of the H$_2$O and CO$_2$ densities, independent of heliocentric distance, subsolar latitude, and the location of the Rosetta spacecraft. We also showed that a fraction of the CO$_2$ is likely associated to the outgassing of H$_2$O. The fitted correlation parameters have been compared to relevant TPD laboratory experiments from the literature. In conclusion, the following key results were obtained:

\begin{itemize}
\item[--] The results for comet 67P show that highly volatile molecules are simultaneously associated with the two major species, H$_2$O and CO$_2$, although the proportions vary depending on the species.
\item[--] No ice phase dominated by highly volatile species, which sublimate at temperatures well below $\SI{70}{\kelvin}$, is required to explain the measured data. A large fraction of these ices have most likely been lost from 67P or they were never part of the comet to begin with. Loss may already occur at even lower temperatures if ample time is available, i.e., internal heating up to  $\SI{\sim70}{\kelvin}$ may not be required to explain the findings at 67P \citep[cf.][]{Parhi2023}. The absence of low temperature ices may also relax the requirements for a cryogenic surface sample return mission to a JFC.
\item[--] TPD experiments from the laboratory show similar behavior as observed at 67P: gases of highly volatile species are trapped in different proportions in or below both H$_2$O and CO$_2$ \citep{Kouchi1995,Notesco2005,Gudipati2023}. This suggests the inheritance of amorphous ices, which trapped these highly volatile species well before the formation of the solar system. For obvious reasons, TPD experiments often employ a single temperature to freeze-out the different species together, e.g., $\SI{10}{\kelvin}$ by \cite{Kouchi1995} and \cite{Gudipati2023}. This may not reflect the formation conditions for the ices incorporated into the comet. Also, in these experiments the samples are heated from the bottom, unlike the situation at a comet, where the ices are heated from the top by solar insolation and erosion and thermal skin depths can be comparable close to the sun. Furthermore, other types of ices cannot be excluded, provided similar behavior can be observed. Additional laboratory and modeling work is required.
\item[--] The presented results are also in line with thermal modeling of comets and KBOs, which shows that significant amounts of ices dominated by highly volatile species, such as CO, may have been lost from these small icy bodies \citep{DeSanctis2001,Parhi2023}. This loss may even be enhanced for a JFC, like 67P, for which internal heating up to $\SI{70}{\kelvin}$ is expected to have taken place during the Centaur stage \citep{Guilbert2016}. Indeed, when comparing families of comets, somewhat elevated CO/H$_2$O ratios may be present in OCCs compared to JFCs \citep{AHearn2012}. ESA's Comet Interceptor mission, which is designed to visit a DNC, will further shed light on the loss of low temperature ice phases \citep{Snodgrass2019}, keeping in mind that also DNCs may have their surface layer substantially processed before reaching the inner solar system \citep{DeSanctis2001,Maggiolo2020,Harrington-Pinto2022,Davidsson2023}.
\item[--] The exact amounts of CH$_4$, CO, O$_2$, N$_2$, etc. that were lost due to thermal and collisional heating \citep{Davidsson2021a,Davidsson2023,Jutzi2020,Parhi2023}, furthermore, depend on the original trapping temperature of the ices incorporated into the comet. Due to the evolutionary loss of the ice phases dominated by highly volatile species, it is very challenging to recover this temperature. A single temperature at which all species freeze out together is unlikely. 
\item[--] Data obtained early in the mission are distinctly different. This was a time when the H$_2$O outgassing was dominated by the evolutionary processed northern hemisphere but CO$_2$ originated from the south. Our results indicate that H$_2$O kept CO$_2$ and highly volatile molecules such as CH$_4$, C$_2$H$_6$, C$_3$H$_8$, O$_2$, and H$_2$S trapped, even if transported as icy grains from the southern to the northern hemisphere where the water sublimation then occurred. The CO$_2$-dominated phase, on the other hand, was lost from such grains. This is in line with numerical simulations by \cite{Davidsson2021a}. 
\item[--] CO$_2$ outgassing during the early mission phase was observed predominantly from the southern hemisphere which was in winter during that time. Our results indicate that the CO$_2$ measured during that time was  processed. A possible explanation is seasonal frost, i.e., CO$_2$, which sublimated during the outbound journey of the previous apparition but refroze before leaving the nucleus. In this process the trapped (highly) volatiles may have been lost. Seasonal CO$_2$ frost has indeed been observed on the surface of 67P's southern hemisphere \citep{Filacchione2016}. We derive an order of magnitude of the seasonally inherited amount of CO$_2$ frost of one percent relative to the total CO$_2$ outgassing.
\item[--] As opposed to the molecules discussed above, the CO abundance was elevated early in the mission. This remains poorly understood because other highly volatile molecules, with comparable sublimation temperatures, exhibit the opposite behavior. For instance, CH$_4$ was depleted in CO$_2$ during the same time period. Interestingly, similarly elevated CO activity, at comparable heliocentric distances during the inbound part of the orbit, has already been observed at comet Hale-Bopp \citep{Biver2002a}. Seasonal CO frost, carried over from the last orbit similar to CO$_2$, would be able to explain these observations. The low temperature required to freeze out CO, however, counters that argument.
\item[--] Additional laboratory experiments are required. It is key that these experiments include both H$_2$O and CO$_2$ in variable proportions, together with the species to be studied, e.g., N$_2$, CO, CH$_4$, etc. The relative proportions of the latter should be similar to the ones found in comets (e.g., 67P in Table~\ref{tab:results}), with highly volatile species being present in trace amounts, maybe somewhat enhanced to account for the loss of the low-temperature ices over the lifetime of the comet (cf. Table~\ref{tab:lost}). We do not know the detailed formation of the main ices incorporated in the nucleus. If the H$_2$O and CO$_2$ ices formed separately, for instance in a layered mantle, similar to the so-called Greenberg particles \citep{Greenberg1982}, then the trapping (or chemical formation) of trace species may be subject to vastly varying CO$_2$/H$_2$O ratios. Laboratory measurements should therefore also cover CO$_2$/H$_2$O ratios that are higher than what is commonly observed in comets. Furthermore, measurements with varying trapping temperatures are required to investigate whether the formation temperature of the comet can be recovered from the relative partitioning of highly volatile species in H$_2$O and CO$_2$ ices, given that the ice-phases dominated by highly volatile species will be mostly lost afterwards. Last but not least, also frost cycles of H$_2$O, CO$_2$, and possibly CO ices containing impurities ought to be investigated further.
\end{itemize}

\section*{Acknowledgements}
We gratefully acknowledge the work of the many engineers, technicians and scientists involved in the Rosetta mission and in the ROSINA instrument in particular. Without their contributions, ROSINA would not have produced such outstanding results. Rosetta is an ESA mission with contributions from its member states and NASA. Work by M.R., N.H., and D.R.M. was funded by the Canton of Bern and the Swiss National Science Foundation (200020\_207312). M.R.C and Y.S. were funded by NASA grants 80NSSC18K1280 and 80NSSC22K1064, respectively. M.S.G. thanks NASA's DDAP program for funding and his part of the work was carried out at the Jet Propulsion Laboratory, California Institute of Technology, under a contract with the National Aeronautics and Space Administration (80NM0018D0004). N.F.W.L. and K.A.K. acknowledge support from the Swiss National Science Foundation (SNSF) Ambizione grant 193453 and NCCR PlanetS. S.F.W. acknowledges the financial support of the SNSF Eccellenza Professorial Fellowship (PCEFP2\_181150).\\
We thank the referee, Gianrico Filacchione, for the constructive feedback that enabled us to significantly improve the quality of our manuscript.

\section*{Data Availability}
All Rosetta/ROSINA data is available through the NASA Planetary Data System (PDS) and ESA's Planetary Science Archive (PSA).



\bibliographystyle{mnras}
\bibliography{References} 




\newpage
\appendix

\section{Details on ROSINA DFMS \& COPS}\label{app:ROSINA}
ROSINA was composed of three instruments \citep{Balsiger2007}, the Double Focussing Mass Spectrometer (DFMS), the Reflectron-type Time-of-Flight mass spectrometer (RTOF), and the COmet Pressure Sensor (COPS).  All three sensors were operated through the Data Processing Unit (DPU). The data used in this work is based on the combination of relative gas abundances measured with DFMS and the total gas densities obtained with COPS. Both sensors are described briefly in following.

COPS contained two pressure gauges out of which the Nude Gauge (NG) forms the baseline of this work. The NG was an extractor-type ionization gauge, designed to measure the local gas density surrounding the Rosetta spacecraft. Neutral gas was bombarded by $\SI{150}{\electronvolt}$ electrons inside an open ionization volume surrounded by grids to trap ionizing electrons and guiding freshly formed ions towards the collector electrode. Both grids were very coarse to minimize interaction with the cometary neutral gas flow. At the collector electrode, the current of ions was measured by a highly sensitive electrometer. The ratio of the measured ion current to the regulated electron emission current, nominally set to $\SI{100}{\micro\ampere}$, is proportional to the neutral gas density inside the ionization volume after applying the lab-calibrated instrument sensitivity factors \citep{Graf2004}.

The DFMS was a mass spectrometer in Nier-Johnson design with a mass resolution of $m/\Delta m=3000$ for the full width at 1\% peak height on mass/charge $\SI{28}{\atomicmassunit\per\elementarycharge}$. Neutral gas entering the ion source was ionized by $\SI{45}{\electronvolt}$ electrons from a hot filament emitting a current of $\SI{200}{\micro\ampere}$. The newly formed ions were accelerated and then deflected first by $\SI{90}{\degree}$ in an electrostatic analyzer and then by $\SI{60}{\degree}$  in the field of a permanent magnet. The offset potential of the ion source ensured that external coma ions would not have a suitable energy to pass through this analyzer section. Afterwards, suitable quadrupole electric fields widened up the ion beam, which was hence further separated by mass/charge, onto the Micro Channel Plates (MCP) located on top of a position-sensitive Linear Electron Detector Array (LEDA) with two parallel rows of 512 pixels each \citep{Nevejans2002}.

For a selected set of ion optical potentials only a limited mass/charge range, distributed across the pixels, could be measured (e.g., $\pm\SI{0.25}{\atomicmassunit\per\elementarycharge}$ at $\SI{28}{\atomicmassunit\per\elementarycharge}$). Therefore, the voltages in DFMS had to be adjusted for each mass/charge to be measured. Obtaining a single spectrum took about  $\SI{30}{\second}$, which included some $\SI{10}{\second}$ for setting and settling of the voltages followed by  $\SI{20}{\second}$ of signal integration. A mass scan covering each integer mass/charge from 13 to $\SI{100}{\atomicmassunit\per\elementarycharge}$ therefore lasted about  $\SI{45}{\minute}$. 

If peaks in the MCP/LEDA spectra were overlapping due to close mass/charge, each peak was fitted by the sum of two Gaussians, one about 10\% of the amplitude of the other but 2 to 3 times wider. One example is N$_2$ which is found close to CO, the former a minor volatile and the latter a major cometary parent species \citep{Rubin2015b}. Furthermore, since DFMS measured the mass lines in sequence, the obtained signals of the individual species were linearly interpolated in time such that ratios of different volatile species could be derived.

The ionization process did not only produce parent ions but also break up the molecules. These so-called fragmentation patterns depend on the molecule and the energy of the ionizing electron and had to be taken into account when deriving relative abundances. This includes the subtraction of the signal from higher mass parent molecules, e.g., the electron impact dissociation contribution from $\mathrm{CO}_2+\mathrm{e}^-\rightarrow \mathrm{CO}^++\mathrm{O}+2\mathrm{e}^-$ had to be subtracted from the measured CO$^+$ signal. Once fragmentation has been taken care of, the species-dependent sensitivity factors were applied, taking into account that the detector was operated in analogue mode and had hence a different yield for incident ions. Furthermore, also the transmission through the instrument and the ionization cross section differed among the molecules. Finally, the relative abundances of the coma gases were obtained and, by scaling to the total density measured with COPS, the absolute densities of these species were retrieved. More details on these measurements, how they have been processed, and the associated calibration factors can be found in \cite{Rubin2019a}.

\section{Figure collection of other highly volatile molecules}

\begin{center}
\begin{figure*}
\includegraphics[width=1.0\textwidth]{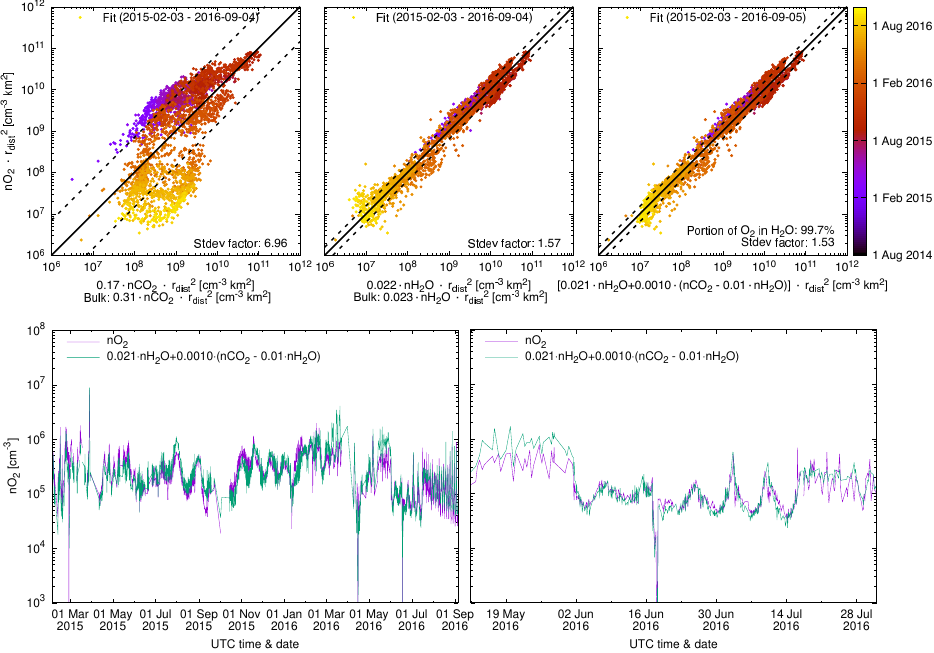}
\caption{Same as Fig.~\ref{fig:CH4} but for O$_2$ instead of CH$_4$.}
\label{fig:O2}
\end{figure*}
\end{center}

\begin{center}
\begin{figure*}
\includegraphics[width=1.0\textwidth]{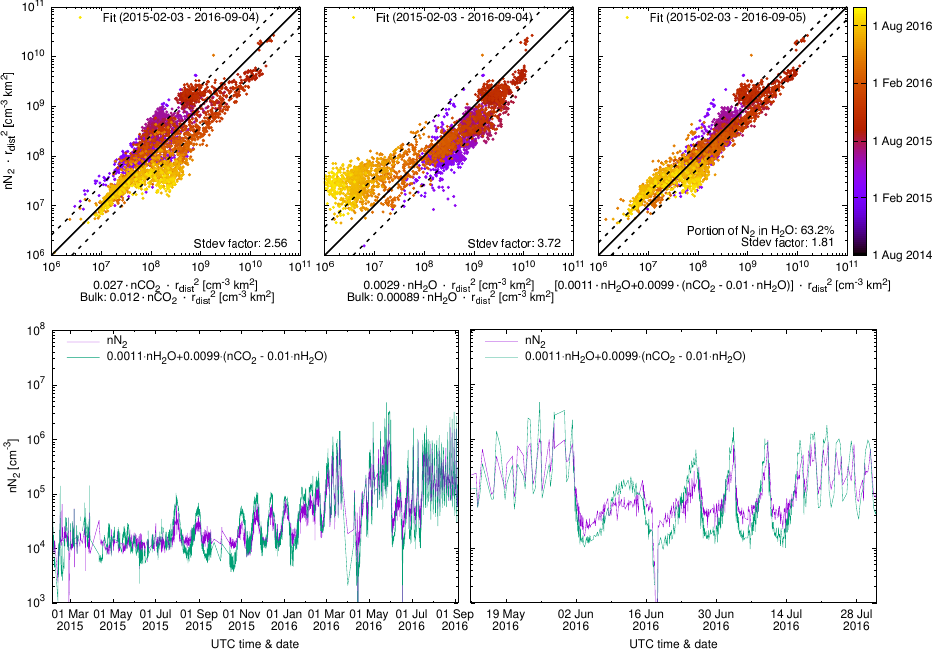}
\caption{Same as Fig.~\ref{fig:CH4} but for N$_2$ instead of CH$_4$.}
\label{fig:N2}
\end{figure*}
\end{center}

\begin{center}
\begin{figure*}
\includegraphics[width=1.0\textwidth]{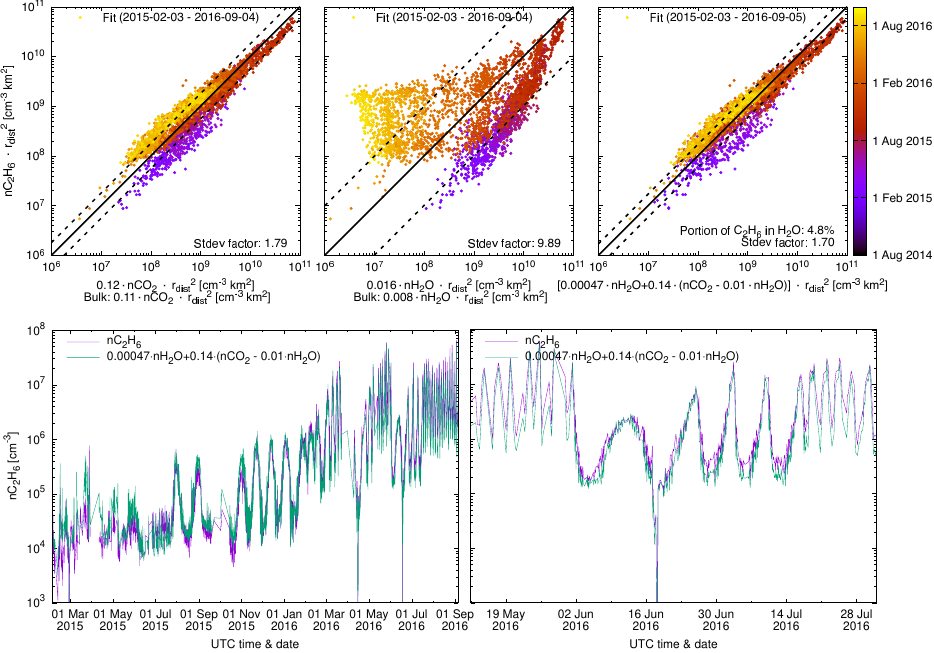}
\caption{Same as Fig.~\ref{fig:CH4} but for C$_2$H$_6$ instead of CH$_4$.}
\label{fig:C2H6}
\end{figure*}
\end{center}

\begin{center}
\begin{figure*}
\includegraphics[width=1.0\textwidth]{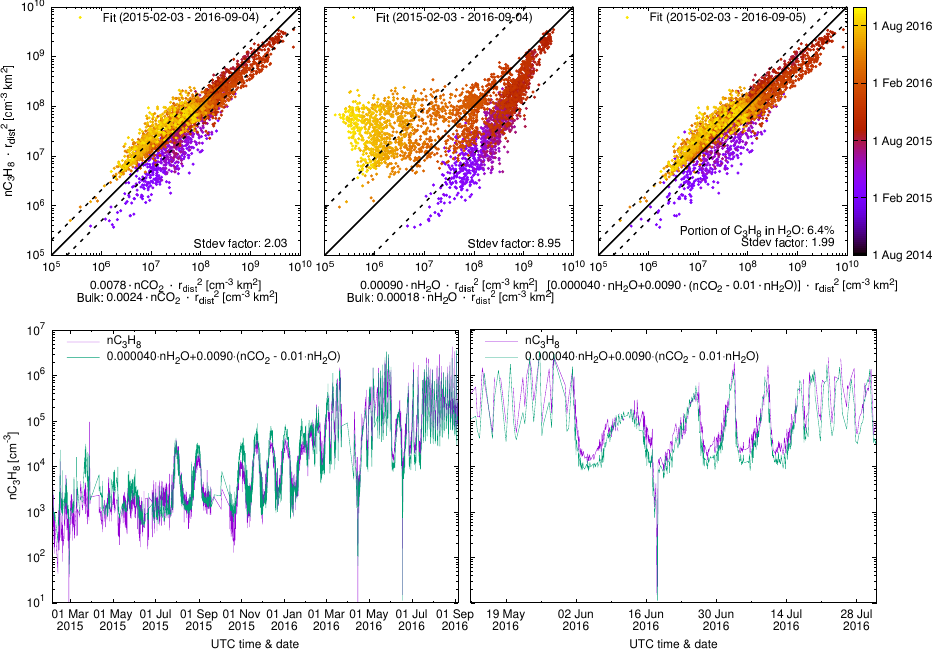}
\caption{Same as Fig.~\ref{fig:CH4} but for C$_3$H$_8$ instead of CH$_4$.}
\label{fig:C3H8}
\end{figure*}
\end{center}

\begin{center}
\begin{figure*}
\includegraphics[width=1.0\textwidth]{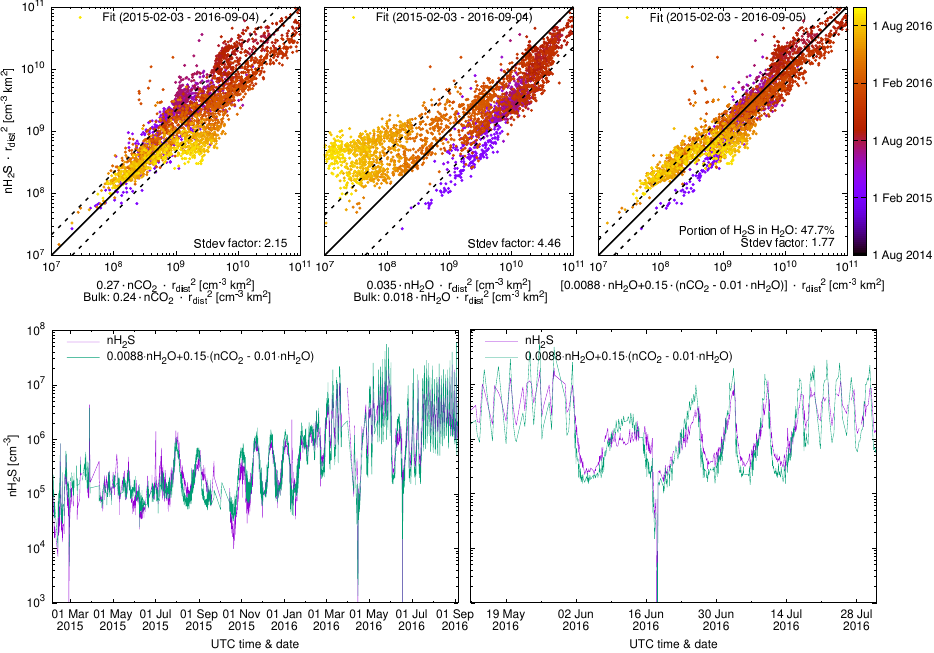}
\caption{Same as Fig.~\ref{fig:CH4} but for H$_2$S instead of CH$_4$.}
\label{fig:H2S}
\end{figure*}
\end{center}



\bsp	
\label{lastpage}
\end{document}